\documentclass[10pt,journal,compsoc]{IEEEtran}
\usepackage{amsmath,amsfonts}
\usepackage{algorithm}
\usepackage{array}
\usepackage[caption=false,font=normalsize,labelfont=sf,textfont=sf]{subfig}
\usepackage{textcomp}
\usepackage{stfloats}
\usepackage{url}
\usepackage{verbatim}
\usepackage{graphicx}
\usepackage[nocompress]{cite}
\usepackage{url}
\usepackage{amsmath, amssymb}

\usepackage{comment}
\usepackage{enumitem} %
\usepackage{booktabs}
\usepackage{multirow}
\usepackage{nicefrac}
\usepackage[normalem]{ulem}
\useunder{\uline}{\ul}{}
\usepackage{array}
\usepackage{tikz}
\newcolumntype{P}[1]{>{\centering\arraybackslash}p{#1}}
\newcolumntype{M}[1]{>{\centering\arraybackslash}m{#1}}
\usepackage{pifont}%
\usepackage{adjustbox}
\usepackage{makecell}
\usepackage{capt-of}
\usepackage{balance} %

\usepackage{tabularx}
\usepackage{siunitx}
\usepackage{graphicx}
\usepackage{pgfplots}

\usepackage{algorithm}
\usepackage{algpseudocode}

\usepackage{colortbl} %
\usepackage{amssymb}

\usepackage{pdfpages}

\newcommand{\simdset}{SimSPAD }
\newcommand{\tldata}{HS-ESPAD }
\newcommand{\supp}{suppl. mat. }

\usepackage{color} 
\newcommand{\changes}[1]{\textcolor{black}{#1}}
\newcommand{\rebuttal}[1]{\textcolor{black}{#1}}

\newcolumntype{C}{>{\centering\arraybackslash}X}

\newcommand{\Sec}{Section~}
\newcommand{\Fig}{Fig.~}
\newcommand{\Tab}{Table~}
\newcommand{\Alg}{Algo.~}
\newcommand{\Eq}{Equation.~}

\definecolor{somegray}{rgb}{0.5, 0.5, 0.5}
\newcommand{\darkgrayed}[1]{\textcolor{somegray}{#1}}

\newcommand{\insertwatermarkabove}{%
  \vspace*{-3em}
  \begin{center}

    \large\darkgrayed{This paper has been accepted for publication at the \\
    IEEE Transactions on Pattern Analysis and Machine Intelligence, 2025. \copyright~IEEE}
  \end{center}
  \vspace{1em}
}

\makeatletter
\pretocmd{\@maketitle}{\insertwatermarkabove}{}{}
\makeatother

\ifCLASSOPTIONcompsoc
  \usepackage[nocompress]{cite}
\else
  \usepackage{cite}
\fi
\def\MYTITLE{Event Cameras Meet SPADs \\ for High-Speed, Low-Bandwidth Imaging}
\title{\MYTITLE}

\author{Manasi Muglikar \textsuperscript{1}
\qquad
Siddharth Somasundaram \textsuperscript{2}
\qquad 
Akshat Dave \textsuperscript{2}
\qquad
Edoardo Charbon \textsuperscript{3}
\qquad \\
Ramesh Raskar \textsuperscript{2}
\qquad
Davide Scaramuzza \textsuperscript{1} \\[6pt]
\,
\thanks{ 
    This work was supported by the European Research Council (ERC) under grant agreement No. 864042 (AGILEFLIGHT).\\
\textsuperscript{1}Robotics and Perception Group, University of Zurich, Switzerland \\
\textsuperscript{2} Camera Culture Group, MIT Media Lab, USA \\
\textsuperscript{3} EPFL, Lausanne, Switzerland
}
}

\IEEEtitleabstractindextext{%
\begin{abstract}
    Traditional cameras face a trade-off between low-light performance and high-speed imaging: longer exposure times to capture sufficient light results in motion blur, whereas shorter exposures result in Poisson-corrupted noisy images. 
    While burst photography techniques help mitigate this tradeoff, conventional cameras are fundamentally limited in their sensor noise characteristics. 
    Event cameras and single-photon avalanche diode (SPAD) sensors have emerged as promising alternatives to conventional cameras due to their desirable properties. 
    SPADs are capable of single-photon sensitivity with microsecond temporal resolution, and event cameras can measure brightness changes up to $1$ MHz with low bandwidth requirements. 
    We show that these properties are complementary, and can help achieve low-light, high-speed image reconstruction with low bandwidth requirements.  
    We introduce a sensor fusion framework to combine SPADs with event cameras to improve the reconstruction of high-speed, low-light scenes while reducing the high bandwidth cost associated with using every SPAD frame. 
    Our evaluation, on both synthetic and real sensor data, demonstrates significant enhancements ($>5$ dB PSNR) in reconstructing low-light scenes at high temporal resolution ($100$ kHz)  compared to conventional cameras. Event-SPAD fusion shows great promise for real-world applications, such as robotics or medical imaging.        
\end{abstract}

}

\newcommand\MYhyperrefoptions{bookmarks=true,bookmarksnumbered=true,
pdfpagemode={UseOutlines},plainpages=false,pdfpagelabels=true,
colorlinks=true,breaklinks=true,
pdftitle={\MYTITLE},%
pdfsubject={Robotics, Computer Vision},%
pdfauthor={M. Muglikar, S. Somasundaram, A. Dave, E. Charbon, R. Raskar, D. Scaramuzza},%
pdfkeywords={Event Cameras, SPADs, Image reconstruction, Low Light imaging}}%
\usepackage[\MYhyperrefoptions,pdftex]{hyperref}

\begin{document}

\maketitle
\IEEEdisplaynontitleabstractindextext
\IEEEpeerreviewmaketitle

\section{Introduction}
High-speed imaging is crucial for diverse applications ranging from safe autonomous navigation to understanding biological tissue dynamics. Attaining high-speed imaging with conventional CMOS and CCD sensors suffers from the challenges of \textit{low signal-to-noise ratio (SNR)} and \textit{high bandwidth}: Low SNR because sampling the scene intensity at a low exposure leads to high noise and high bandwidth because high sampling rate results in a substantial amount of data. Enabling high-speed imaging with adequate SNR and manageable bandwidth demands developing new imaging techniques and sensor architectures beyond conventional cameras.

\begin{figure*}[!t]
    \centering
    \includegraphics[trim={0cm 8.5cm 0cm 0cm},clip,width=\linewidth]{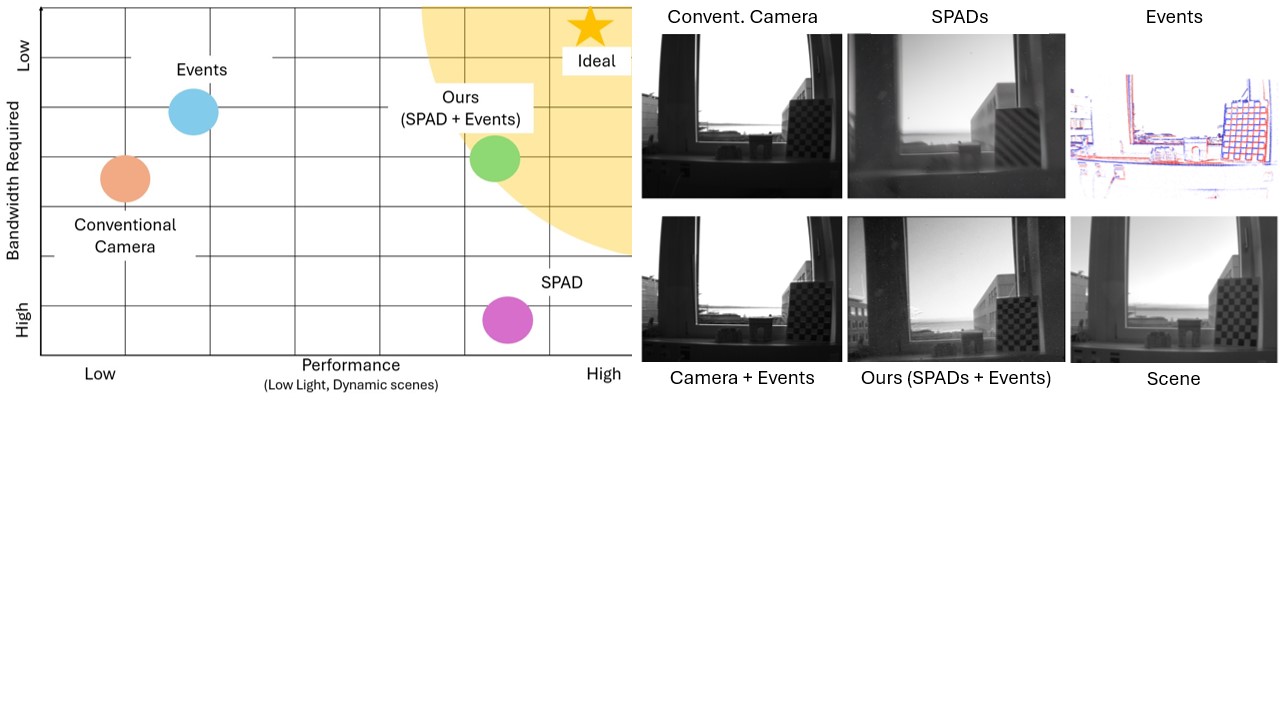}
    \vspace{-1em}
    \caption{\textbf{Towards low-light, high-speed, and low bandwidth imaging by combining event cameras and SPADs.} (Left)Event cameras are able to operate at high speeds with low bandwidth, while SPAD cameras are highly sensitive in low-light conditions. We leverage the complementary properties of these two cameras for high-quality image reconstruction under challenging conditions.
    (Right) Visualliy comparing conventional camera, SPAD camera and event camera in HDR scenes.
    Our method combines events with SPADs resulting in highest quality image reconstructions of both dark indoor region and bright outdoor scene, compared to all sensors while requiring lower bandwidth than SPAD-only methods.} 
    \vspace{-1em}
    \label{fig:eyecatch}
\end{figure*}

We focus on two rapidly advancing sensor technologies aimed at passive high-speed imaging: single photon avalanche diodes (SPADs) and event sensors. These sensors address the challenges in high-speed imaging from different perspectives. SPADs directly convert single photon incidences into counts, eliminating read noise and achieving better SNR than conventional cameras. SPADs can measure scene intensity as 1-bit frames at rates up to 100 kHz, but these bit frames require substantial bandwidth. Conversely, event cameras asynchronously register sparse intensity changes exceeding a certain threshold. As a result, event cameras have much lower bandwidth requirements, but recovering the entire high-fidelity intensity image from sparse events is challenging. Thus, SPADs and event cameras lie at opposite ends of a \textbf{bandwith-performance tradeoff} depicted in  \Fig \ref{fig:eyecatch}.

\global\long\def\figHeight{5cm}

\textbf{Our key insight} is that these complementary capabilities of SPAD and event cameras can be combined to achieve high imaging performance and low bandwidth jointly. We demonstrate that SPADs and events provide superior SNR under a low-light flux regime than conventional cameras. 
We propose the \textbf{first approach} to combine events and SPAD frames for high-speed low-bandwidth imaging. The main idea of our proposed method is to first read aggregated SPAD frames which reduces bandwidth but introduces motion blur. We then deblur using event frames obtained at higher temporal resolution with minimal additional bandwidth to deblur the aggregated SPAD frames.

Our method for sensor fusion of SPAD and event cameras consists of three components. First, we incorporate the non-linear camera response function of SPADs for a new method for deblurring SPAD frames from events that we term Nonlinear Event Double Integration (NEDI). Second, our sensor fusion strategy uses a Kalman-filter-based approach that fuses the asynchronous events and deblurred SPAD frames to reconstruct HDR images in continuous time. Third, to further tackle bandwidth constraints, we propose an adaptive sampling approach for SPAD frames. The uncertainty estimates from the Kalman filter are leveraged to adaptively trigger SPAD frames once the uncertainty falls below a threshold. 

We evaluate the complementary properties of SPAD and event cameras and our sensor fusion technique on both simulated and real-world scene. For real-world comparisons, we created a dataset with calibrated and aligned SPAD, event and conventional cameras. SPAD-based  Our sensor fusion allows us to achieve high-speed ($10kHz$), low light ($> \SI{10}{Lux}$) image reconstruction compared to event and conventional frame fusion methods which can only achieve $5000fps$ at $> \SI{100}{Lux}$ \cite{Tulyakov22cvpr}. Moreover, our adaptive approach not only reduces the bandwidth requirements of SPAD frames by 4$\times$, but also achieves 5 dB gain in PSNR compared to conventional camera-based approaches. We achieve better performance than SPAD-only baselines such as Quanta Burst Photography \cite{Ma20TOG}, while reducing the bandwidth and power by $1000 \times$. 

\textbf{Contributions.} We summarize our key contributions as follows:
\begin{itemize}
    \item We demonstrate the complementary properties of SPAD and event cameras for high-speed imaging on a novel real-world dataset
    \item We demonstrate that there exist low light flux regimes where both SPAD and event cameras have better SNR than conventional cameras 
    \item We propose the first sensor fusion method to combine SPAD frames and events for high-speed imaging under low light and bandwidth constraints.
    \item Our proposed method achieves better imaging performance conventional than state-of-the art conventional frame-event fusion, events-only and SPAD-only techniques while significantly reducing bandwidth over SPAD-only techniques by $1000 \times$ and over the conventional frame-event fusion by $2 \times$. %
\end{itemize}

\vspace{-1em}
\section{Related Work}
\label{sec:relwork}
\textbf{HDR imaging with frames}
Burst denoising is a popular technique to capture low light scenes by merging and denoising multiple frames \cite{Hasinoff16tog, Liba19tog}.
These methods often rely on motion estimation and alignment to merge the frames. Deep learning-based approaches were also proposed to address automatic alignment \cite{Godard18ECCV, Mildenhall18CVPR}.
While these methods focus on obtaining single clean image from multiple noisy frames, other works such as \cite{Monakhova22CVPR} focus on video denoising.
However, at extremely low light, conventional cameras suffer from a low signal to noise (SNR) ratio, which makes it difficult to capture the scene.
This is primarily because the pixel electronics noise dominates the signal at low light levels \cite{Ingle21CVPR}.
It is precisely this limitation that has motivated the development of SPADs for high dynamic range imaging.

\noindent \textbf{Low bandwidth imaging with event cameras}
Event cameras provide sparse relative intensity changes with a very low bandwidth.
Extracting the full intensity from these sparse measurements was shown in several works \cite{Scheerlinck18accv,Cook2011, Kim14BMVC, Bardow16CVPR, Munda18IJCV}.
Significant advancement was done by learning to reconstruct images from events \cite{Rebecq2021, Barua16WACV}.
These reconstructed images inherited the HDR property of event cameras.
However, in the absence of contrast or relative scene motion, these methods do not produce any meaningful information.
Several methods were then proposed to combine frames with events to obtain the low frequency details from the frames and the high frequency illumination changes from events \cite{pan19cvpr, wang21iccv, Tulyakov21CVPR, Tulyakov22cvpr, messikommer22cvprw, Wang24pami}.
\rebuttal{Recent approaches have focused on combining frames and events to obtain high-quality images in low light.
Liu et al. \cite{Liu24cvpr} propose to reconstruct images from events using the illumination characteristics of event camera to train an image reconstruction module. 
As obtaining paired ground truth for such tasks is quite challenging, this method uses day-time images as ground truth to train this network.
Song et al. \cite{Song20eccv} propose an unsupervised domain adaptation method for events which prioritizes reconstruction of specific day-light features of the scene.
Liang et al. \cite{liang24cvpr} propose a novel dataset with synchronized events and low-light frames to train an event-guided image enhancement network. 
The aligned ground truth was achieved by mounting the setup on a robotic arm and moving it to capture the same scene with in different lighting.
Although this dataset paves the way for training high-quality image enhancement networks, it is limited to static scenes and requires a complex setup to capture the data.
Furthermore, the inherent limitation of frame-based sensors makes these fusion sensors unusable in low-light HDR scenes.}

\noindent \textbf{HDR imaging with SPADs}
While SPADs have been primarily explored with active sensing applications such as fluorescence microscopy \cite{Perenzoni15ISSCC}, non line of sight imaging, and time of flight imaging \cite{Chao19JSSC, Lindell20natcomm, rapp20spm, Yoshida20eetimes}, only recently have SPADs been utilized as passive imaging devices.
\cite{Ingle21CVPR} proposed a method to capture light intensity by measuring the time between two photons, enabling the capture of high dynamic range scenes.
To combat motion blur, \cite{Ma20TOG} proposed a quanta burst photography (QBP) method to capture high dynamic range scenes with SPADs and correct for motion blur.
Similar to burst photography, this approach focused on obtaining single clean image from multiple binary frames.
Moreover, these methods require each every binary frames of the SPAD sensor to be transmitted, which results in high bandwidth requirements.
Instead, we propose to combine SPADs with events to address the motion blur artifacts.
Our Kalman filter-based approach enables adaptive sampling of SPADs, thus addressing the bandwidth limitations.

\vspace{-1em}
\section{Sensor Characteristics}
\label{sec:sensors}
\begin{figure*}
    \centering
    \includegraphics[trim={0cm 11cm 0cm 0cm},clip,width=\linewidth]{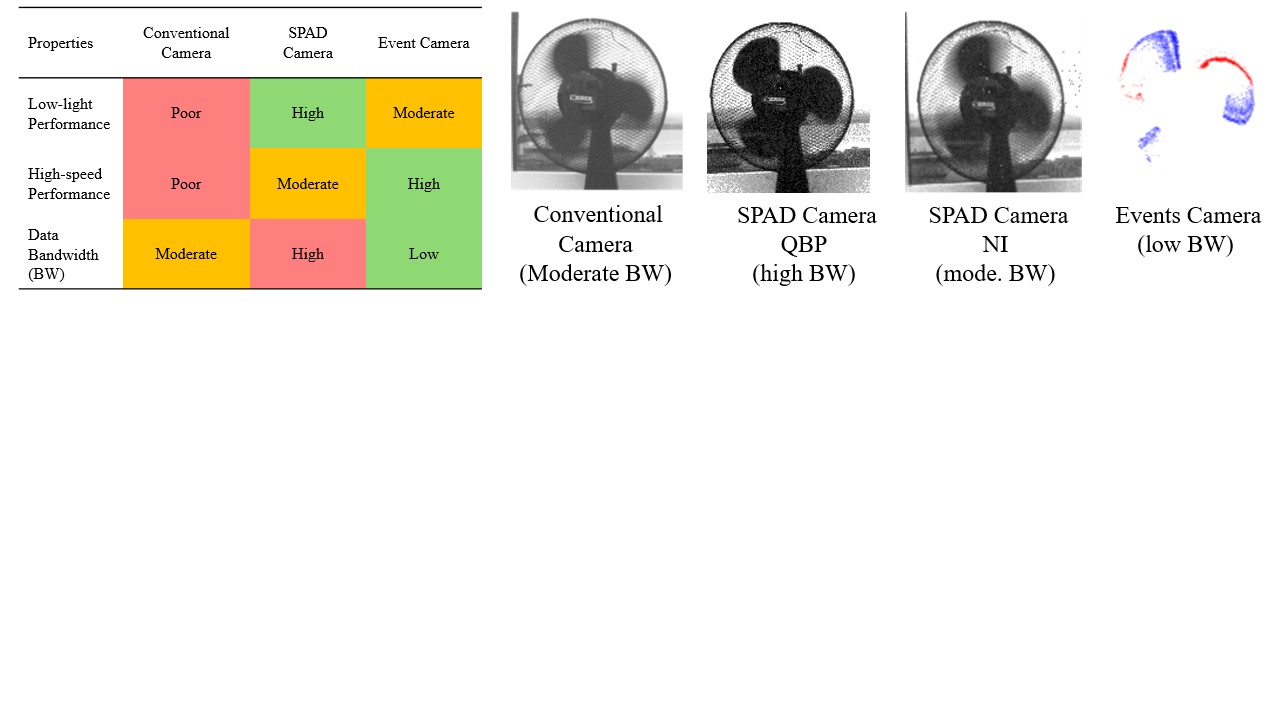}
    \vspace{-1em}
    \caption{\textbf{Comparing sensor characteristics.} Standard cameras require sufficient accumulation of photons, which is impractical for low-light and fast-moving scenes. SPADs measure light through avalanching to enable single-photon sensitivity and high temporal precision, which are useful for low-light and high-speed respectively. Event cameras, by only recording changes in brightness, can operate at high speeds and low bandwidth.} 
    \label{tab:sensor_comparison}
    \vspace{-1em}
\end{figure*}

In this section, we compare the key properties of conventional, SPAD and event cameras through simulated and captured measurements. 
The comparison highlights the inherent trade-offs and strengths of each sensor type, emphasizing that no single imaging modality excels universally across all desired properties.  
The comparison is summarized in \Fig \ref{tab:sensor_comparison}.
We approach this comparison from the perspective of low-light imaging, high-speed imaging and data bandwidth, as these are the key properties that we aim to address in our work.
With this comparison, we build intuition towards our solution for combining the strengths of SPAD and event cameras towards performing high-speed low bandwidth low-light imaging.
\vspace{-1em}
\subsection{Overview}
Conventional cameras and SPADs operate in a fundamentally different manner compared to event cameras in terms of how they capture scene information. 
Conventional cameras and SPADs synchronously sample the scene at fixed time intervals, with each sample containing the raw intensity values for the entire image frame. 
This results in a sequence of intensity frames that represent the scene over time.
Therefore, for both conventional cameras and SPADs, the image frame quality is only a function of the scene illumination ($\phi$) and the sensor noise.

In contrast, event cameras employ an asynchronous sensing mechanism, capturing information only when a change in intensity occurs within a pixel. 
These changes are thresholded and encoded as binary events, each event containing only the polarity (increase or decrease) of the intensity change, not the absolute intensity value itself. 
As a result, event cameras do not capture full-intensity frames. 
Instead, they accumulate a sparse, asynchronous stream of events that encode the temporal changes in the scene.
Therefore, the quality of the event data is a function of the scene illumination ($\phi$), the illumination change ($\Delta \phi$) and the sensor noise.
We now compare the captured scene information of the three sensors in terms of low-light performance, high-speed performance and data bandwidth.
\vspace{-1em}
\subsection{Low Light Performance}
We study the dynamic range of the three sensors by comparing their image quality in capturing different illumination scenes.
To quantify the dynamic range of each sensor modality, we measure the signal to noise ration (SNR) of the captured sensor data at different illuminations.
The SNR is a measure of the quality of the captured image, with higher SNR indicating better image quality.
For traditional cameras and SPAD, it is possible to quantify the performance in terms of signal to noise(SNR) ratio for  any given illumination level.
The SNR comparison between traditional camera and SPADs was shown in \cite{Ingle21CVPR} as expressed as follows:

\begin{equation}
    \text{SNR}_{\text{camera}}(\Phi) = 
    \begin{cases} 
    10 \log_{10} \left( \frac{\Phi}{q\Phi T + \sigma_f^2} \right), & \Phi < \frac{N_{\text{FWC}}}{qT} \\
    -\infty, & \Phi \geq \frac{N_{\text{FWC}}}{qT}
    \end{cases}
\end{equation}

\begin{align}
    \text{SNR}_\text{SPAD}(\Phi) = & -10 \log_{10} \left( \frac{\Phi_{dark}}{\Phi} + q(1 + q\Phi\tau_p)e^{-q\Phi\tau_d} \right)^2 \nonumber \\
    & + \left( \frac{1 + q\Phi\tau_d}{q\Phi T} \right)^2 + \left( \frac{1 + q\Phi\tau_d}{1 - q^2 \Phi^2 \tau^2} \right)^2
\end{align}

where $\Phi$ is the illumination, $\Phi_{dark}$ is the dark current, $q$ is the charge of an electron, $\tau_p$ is the time constant of the photodetector, $\tau_d$ is the time constant of the detector, $T$ is the integration time and $\tau$ is the time constant of the readout circuit.

Due to the unique asynchronous and sparse nature of event camera, quantifying it's performance under different illumination and scene motion is challenging.
SNR in the context of event camera depends not only on the illumination but also on the illumination change and the scene motion.
We therefore simplify the problem as follows:
Consider the scenario where the illumination is given by $\Phi$ and the illumination change is given by $\Delta \Phi$.
The SNR for event camera is given by:
\begin{equation}
    \text{SNR}_{\text{events}}(\Phi, \Delta \Phi) = 10 \log_{10} \frac{\left( P_e * \frac{\Delta \Phi}{C} \right)}{\text{N}(\phi)}
\end{equation}
where $P_e(\Phi, \Delta \Phi)$ is by the event trigger probability  $C$ is the contrast threshold  and $N(\phi)$ the static noise.
Therefore, given event probability, contrast threshold and noise, we can compare the SNR of event camera with traditional cameras and SPADs.
Estimating these values is challenging and is an open problem in the field of event camera research (\cite{Graca23CVPR, Gao23aim}).
In this paper, we resort to empirical measured values of event probability and noise.
Details about calculating SNR for event camera are provided in the \supp

\input{floats/fig_real_sensor_compairson.tex}
In \Fig \ref{fig:snr}, we compare the SNR of all $3$ sensors: frame, SPAD and event camera with increasing illumination.
For event cameras, we show the SNR for $2$ illumination changes: $30\%$ and $100\%$ given a contrast threshold of $30\%$.
We make three main observations from this plot:
\begin{itemize}
    \item SPADs have higher SNR than cameras at both low flux and high flux regime.
    \item More illumination change, better the SNR for events.
    \item Comparing the three sensors, we can see that SPADs and events have a better SNR at lower illumination (with SPADs being more sensitive at lower illumination). 
\end{itemize}
We also show the effect of this using real sensor data in \Fig \ref{fig:real_illum_viz}.
At extremely low illumination ($1Lux$), the frames and events are unable to capture the contrast edges of the star, whereas SPADs can capture these edges.
As the illumination increases to $10Lux$, the events and SPADs both are able to capture the contrast edges of the star being rotated, where conventional cameras struggle.

\vspace{-1em}
\subsection{High-speed performance}
Scene dynamics, characterized by the rate of change in the scene over time, is another crucial factor influencing sensor performance. For slow scene dynamics, where changes occur gradually, event cameras perform similarly to periodic sampling sensors like conventional frames and SPADs, with the mean square error (MSE) of event cameras depending primarily on the contrast threshold, while the MSE of frames and SPADs is determined by the frame rate. As SPADs generally operate at higher frame rates, they outperform conventional frames in slow-motion scenarios, albeit at the cost of higher bandwidth requirements.
In contrast, for fast scene dynamics involving near-instantaneous changes ($< \SI{10}{\micro \sec}$), event cameras exhibit a distinct advantage. 
Their MSE is binary, either zero or one, depending solely on the contrast threshold, while the error for frames and SPADs is a function of the frame rate \cite{Censi15ICRA}. 
Consequently, the higher frame rates of SPADs result in significantly lower error compared to conventional frames in fast-motion scenarios.
Event cameras, however, excel in such conditions, offering the best trade-off between error and bandwidth requirements. 
This is exemplified in \Fig \ref{fig:real_illum_viz}, where a rapidly rotating table fan causes motion blur and low contrast in conventional frames, while SPADs, though struggling, can capture some scene structure behind the fan owing to their high dynamic range. 
Event cameras, on the other hand, can effectively capture the high-contrast edges of the fan's motion.

\vspace{-1em}
\subsection{Data bandwidth}
The data bandwidth of a sensor is a measure of the amount of data it generates per unit time.
For conventional cameras, the data bandwidth is determined by the frame rate and the resolution of the sensor.
For SPADs, the data bandwidth is determined by the frame rate, the resolution of the sensor and the dynamic range of the sensor.
For event cameras, the data bandwidth depends on relative scene motion and is measured in terms of event rate.
The event rate is a measure of the number of events generated per unit time.
The event rate is a function of the scene motion, the illumination and the illumination change.
We use the memory as a proxy for the data bandwidth of the sensor.
For the SPAD camera, the average memory required to store the sensor data was $\SI{3.2}{Gb}$.
Traditional frames stored as in raw file format took up $\SI{0.5}{Gb}$ of memory.
For the same scene, the average memory required to store the event camera data was $\SI{0.154}{Gb}$.

\noindent The summary of the comparison is shown in \Tab \ref{tab:sensor_comparison}.
It shows the complementary nature of SPADs and Event cameras which results in a better trade-off between low light performance, high speed performance and data bandwidth.
\vspace{-2em}
\subsection{Outlook}
\begin{table}[t!]
    \centering
    \begin{tabularx}{\linewidth}{|X|X|X|X|X|}
    \toprule
     Property & \multicolumn{2}{c|}{SPADs} & \multicolumn{2}{c|}{Event Camera} \\
     & Current & Future & Current & Future \\ \midrule
     Latency & $\SI{10}{\micro\second}$ (Readout rate) &  $\SI{10}{\nano\second}$ (Deadtime) & $\SI{1}{\micro\second}$ (Refractory period and bus congestion) & $\SI{10}{\nano\second}$ (Refractory period)\\  \midrule
     Low Light Illumination & $>\SI{1}{Lux}$ (Quantum efficiency (0.4), fill factor) & $<\SI{1}{Lux}$ & $>\SI{10}{Lux}$ (Fill factor, dark photon current) & $<\SI{10}{Lux}$  \\  \midrule
     High Illumination & $<10^{9} Lux$ (Soft saturation) &  $<10^{9} Lux$ (Soft saturation)  & $<10^{4} Lux$ (contrast sensitivity ($>15\%$)) & $>10^{12}Lux$ \\ 
    \bottomrule
    \end{tabularx}
    \caption{\textbf{Current and Future Trends in SPADs and Event Cameras} In terms of low-latency, both SPADs and event cameras in future will have similar latency. 
    In the case of illumination in future SPADs will have better low light sensitivity, whereas improvements in event cameras will result in better bright-light sensitivity.}
    \vspace{-3em}
    \label{tab:sensor_comparison_future}
\end{table}
While our evaluation on real data is based on current hardware, here we discuss how the technological trends will change over time and demonstrate that the two sensors will continue to co-exist and exhibit this complementary nature.
These trends are summarized in \Tab \ref{tab:sensor_comparison_future}.
In the case of SPADs, the inherent limitation for frame rate comes from the pixel deadtime which is in the order of nanoseconds and readout time which is in the order of microseconds. With rapid progress towards reducing bandwidth of single-photon imaging through better communication protocols and in-sensor and near-sensor computing \cite{sheehan2022spline,zhang2022first}, readout time of SPADs is projected to reduce. 
In the case of event cameras, the bottleneck for latency comes from the pixel refractory period, which in the end depends on the capacitor discharge time constant and is in the order of microseconds, and the event bus congestion \cite{Yang17TSC}.
In future, with better asynchronous readout protocol, the bus congestion-related latency could be mitigated to a large extend.
Therefore in context of low-latency event cameras, the capacitor used in the circuit will control the latency, which can be easily reduced by having higher refractory current flowing through the circuit \cite{Delbruck21CVPRW} bringing it down to nanoseconds.

In the context of low-light sensitivity, the limitation for event cameras comes from low contrast sensitivity ($>15\%$).
This means that smaller illumination changes cannot be easily detected by event cameras, which is specially important for high-flux scenarios as seen in the SNR plot \ref{fig:snr:illumination}.
Efforts have been made to improve the temporal contrast sensitivity of event cameras, leading to experimental sensors with higher sensitivity of $1\%$\cite{Serrano13JSSC, Yang15JSSC, Moeys18TBCS}.
Another limitation for low-light sensitivity of event cameras is the fill factor (i.e., the ratio of a pixel’s light sensitive area to its total area).
The previous versions of event cameras featured front-side illuminated circuit, they resulted in lower fill factor $< 20\%$, resulting in poor low contrast sensitivity.
Current and future event camera models incorporated Back-Side Illumination (BSI) technology which improved the fill factor significantly ($>70\%$).
In the case of SPADs, the low-light sensitivity is limited by the quantum efficiency of the sensor which is currently upto $0.4$ \cite{Ingle21CVPR}.
Improvements in this direction can result in the highest performance of SPADs in all illumination scenarios.

Specifically for high illumination scenarios, SPADs suffer from soft-saturation (as seen in the SNR plot \ref{fig:snr:illumination}) resulting in poor SNR beyond $10^9 Lux$.
This is an inherent limitation of the sensor characteristic which depends on the pixel deadtime, which leave little room for improvement.
In contrast, for event cameras, the limitation for operation in high flux regime is the contrast sensitivity.
By adopting better capacitor design, this contrast sensitivity can significantly improve the performance for low illumination changes in high flux regime to beyond $>10^9 Lux$ \cite{Delbruck21CVPRW}.
It was shown in \cite{Delbruck21CVPRW}, that refractory period inversely depends on the capacitor current.
Increasing this current, can result in higher contrast sensitivity at brighter illuminations resulting in high response curve for event cameras across all illumination changes (i.e the light blue curve of event camera will go upto dark blue curve in \ref{fig:snr:illumination} for all illumination).
An important thing to note is that for high illumination change, this is already the case for event cameras.

Looking at these trends, it can be seen that while sensor improvements will result in better latency for both event cameras and SPADs, for the case of illumination they will always provide complementary properties with SPADs providing better low-light sensitivity and event cameras having better signal at high illuminations.
This motivates our work to combine the two sensors to address the combined capabilities of each, which marks landmark in this direction, opening new directions for future more sophisticated sensor fusion strategies.
We will now describe our approach to combine the two sensors in the next section.

\section{Methodology}
\label{sec:methodology}
In \Sec \ref{subsec:deblurring}, we describe our non-linear deblurring method to deblur SPAD images using events.
We describe our asynchronous integration of SPADs and event cameras using a Kalman filter \ref{sec:akf}.
Finally, in Sec. \ref{sec:adaptive} we describe our approach for adaptively changing the integration time of SPADs to reduce the bandwidth.
An overview of our approach is shown in \Fig \ref{fig:method}.

\global\long\def\figWidth{\linewidth}
\begin{figure}
    \includegraphics[width=\figWidth]{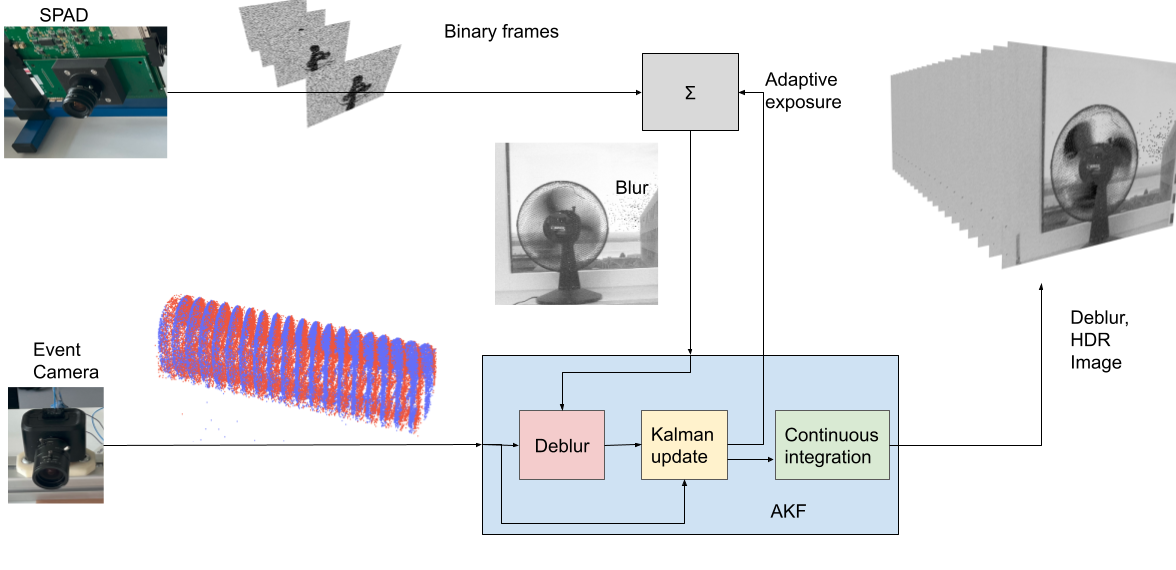}
    \vspace{-3em}
    \caption{\textbf{Overview of our approach.} Our approach operates on the motion blurred images of SPADs. 
    Aligned and synchronized events are then used to deblur the SPAD images.
    The Kalman update uses the deblurred SPAD images and events and generates state estimate and state uncertainty. 
    The uncertainty is used to sample the next SPAD image and state estimate is fused in continuous time to generate the final deblurred HDR images. 
    }
    \label{fig:method}
\end{figure}

\subsection{Deblurring SPAD images with Event Cameras}
\label{subsec:deblurring}
Unlike traditional image sensor, the SPAD camera response function is not linear and therefore the classic Event Double Integration (EDI) method \cite{pan19cvpr} cannot be used here.
The SPAD response function is given by:
\begin{equation}
    \hat{\Phi}_{\text{SPAD}} = \frac{{N}(t)}{qT_{bin}+\tau {N}(t)}
    \label{eq:spad}
\end{equation}
where $N$ is the number of photons detected by the SPAD sensor, $q$ is the quantum efficiency of the SPAD sensor, $T_{bin}$ is the integration time for each binary frame and $\tau$ is the dead time of the SPAD sensor.
Therefore, a blurry image can be expressed as the sum of the true latent image ${N}(t)$ over the period of exposure time $T$ :

\begin{align}
    B_{\text{SPAD}}(t)&= \frac{1}{T} \int_{f-T/2}^{f+T/2} {\Phi}_{\text{SPAD}}(t) dt \\
    &=  \frac{1}{T} \int_{f-T/2}^{f+T/2} \frac{N(t)}{qT_{bin}+\tau N(t)} dt \\
\end{align}
The latent image sequence ${N}(t)$ can be expressed as a function of intensity changes $E(t)$ (obtained from event signal) and the previous latent image ${N}(f)$ as follows:
\begin{equation}
    {N}(t) = {N}(f) \exp\left( c E(t) \right)
\label{eq:latent}
\end{equation}

Substituting $N(t)$ from \Eq \ref{eq:latent}:
\begin{align}
    B_{\text{SPAD}}(t)&= \frac{1}{T} \int_{f-T/2}^{f+T/2} \frac{N(f) \exp\left( c E(t) \right)}{qT_{bin}+\tau N(f)\exp\left( c E(t) \right)}  dt \\
    &= \frac{N(f)}{T} \int_{f-T/2}^{f+T/2} \frac{\exp\left( c E(t) \right)}{qT_{bin}+\tau N(f)\exp\left( c E(t) \right)} dt
    \label{eq:nedi}
\end{align}
Note the difference between this model and the traditional camera model is that while traditional model has a linear relation between the latent image and events, its not the case for SPADs.
Estimating the latent image $N(t)$ from the SPAD image $B_{\text{SPAD}}(t)$ is challenging due to the non-linear relation between the two.
Therefore, we propose to solve this using a optimization framework.
We initialize the latent image  using the linearized form estimated using EDI
We then optimize the latent image using the above equation to minimize the difference between the blurred SPAD image and the estimated blur image produced with a known latent image.
We call this the nonlinear event-based double integral (NEDI) model for SPADs.
The effect of the non-linear relation between the latent image and the SPAD image is shown in \Fig \ref{fig:nedi}.
Using our proposed NEDI model, we can estimate the latent image $\hat{N} (t)$ with sharper edges than the EDI model.

\global\long\def\figWidth{0.45\linewidth}
\begin{figure}[!h]
	\centering
    \setlength{\tabcolsep}{2pt}
	\begin{tabularx}{\linewidth}{
        M{0.2cm}
        M{\figWidth}
        M{\figWidth}}
        & \makecell{EDI\cite{pan19cvpr}}& {\makecell{NEDI (Ours)}} \\
        \rotatebox{90}{\makecell{Fan}}&
        \includegraphics[trim={3.5cm, 5cm, 10cm, 7cm},clip,width=1.0\linewidth]{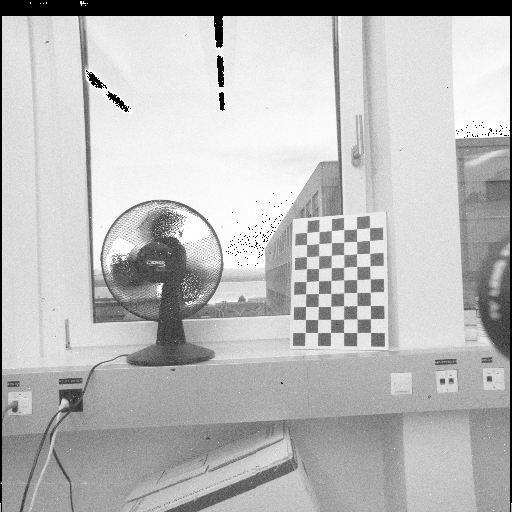} &
        \includegraphics[trim={3.5cm, 5cm, 10cm, 7cm},clip,width=1.0\linewidth]{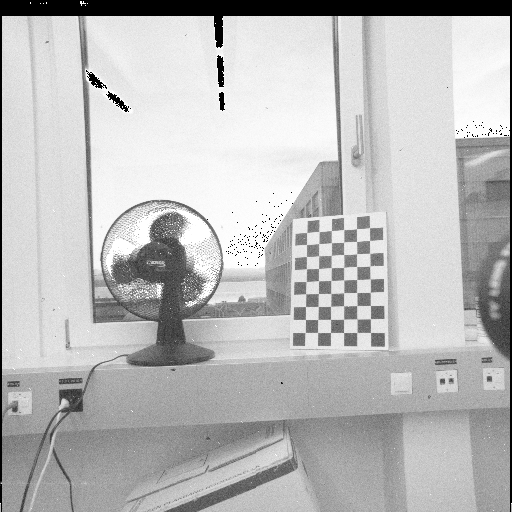}\\
        \rotatebox{90}{\makecell{Zoom}}&
        \includegraphics[trim={4.5cm, 9cm, 11cm, 7cm},clip,width=1.0\linewidth]{images/nonlinear/linear_00001.png.png} &
        \includegraphics[trim={4.5cm, 9cm, 11cm, 7cm},clip,width=1.0\linewidth]{images/nonlinear/nonlinear00001.png.png}\\
    \end{tabularx}
    \caption{\textbf{Non-linear event double integration de-blurring (NEDI).} Our non-linear (NEDI) deblurring approach qualitatively produces sharper reconstructions than EDI\cite{pan19cvpr}. 
    }
    \label{fig:nedi}
    
    \vspace{-1em}
\end{figure}
\vspace{-1em}
\subsection{Asynchronous Kalman Filter}
\label{sec:akf}
We introduce the kalman filter which integrates the uncertainty models of events and SPAD measurements to computer the kalman gain.
Our AKF is inspired by the work of \cite{wang21iccv}.
While the backbone of our AKF remains the same, we extend this approach to include the uncertainty models of SPAD sensors.
We first introduce the uncertainty models used for event camera and SPAD camera in \Sec \ref{sec:eventnoise} and \Sec \ref{sec:spadnoise} respectively.
We then describe the asynchronous integration of SPAD frames using the AKF backbone in \Sec \ref{sec:akf}. 

\subsubsection{Event Camera Uncertainty Model}
\label{sec:eventnoise}
The models for noise in event camera are difficult to develop due to the complex circuit behaviour. 
In \cite{wang21iccv}, they proposed a simple heuristics to model the event noise as additive Gaussian process.
The model considers three types of noise: (a) process noise (b) isolated pixel noise and (c) refractory period noise.
However, their model did not consider the effect of illumination and contrast threshold mismatch on the noise \cite{Graca23CVPR}.
In this paper, we extend their model to include the noise dependence on illumination and contrast threshold mismatch in accordance to circuit noise model.
The noise is modelled as a Gaussian process with variance that grows linearly with time since the last pixel occurred as follows:
\begin{equation}
    Q  = \sum_{i=0}^{\infty} (Q^{shot.} + Q^{isol.} + Q^{ref.} + Q^{thresh.}) \delta (t - t^i)
    \label{eq:ev_noiseco}
\end{equation}

\noindent \textbf{Temporal noise/Shot noise}
This noise models the effect of the photon shot noise which depends on the incident photon flux $\phi$.
The noise is insignificant at high flux but dominates at low flux.
We model this using a function $f (\phi)$ which decreases with increasing flux and drops to zero.
It is modelled as a Gaussian process with variance that grows linearly with time since the last pixel occurred as follows:
\begin{equation}
    Q^{shot} = f (\phi) t^i - t^{i-1}
\end{equation}

\noindent \textbf{Threshold mismatch}
The typical value of contrast sensitivity is about $0.3$.
The uncertainty of the contrast sensitivity is modelled as a Gaussian distribution with variance $\sigma_{\theta} = 0.03$.

\noindent \textbf{Isolated pixel noise/Hot pixels}
Hot pixels tend to occur in isolation and are not correlated with other pixels.
This noise was modelled using a variance as:
\begin{equation}
    Q^{isol.} = \sigma_{iso}^2 (t^i - t^{i-1})
\end{equation}
\noindent \textbf{Refractory period noise}
The refractory period noise is a result of the refractory period $\rho$ of the pixels.
Within this period, no event will be triggered due to circuit limitations.
\begin{equation}
    Q_p^{ref.} =
    \begin{cases}
    $0$      & \text{if $t_p^i-t_p^{i-1}> \rho$} \\
    \rho_{ref}      & \text{otherwise} \\
    \end{cases} 
\end{equation}

\subsubsection{SPAD Camera Uncertainty Model}
\label{sec:spadnoise}
The noise in scene irradiance comes from uncertainty in raw camera response, also known as Camera Response Function (CRF).
\begin{align}
    I_p &= CRF^{-1}\left(F_p\right) + \mu_p, \quad
\mu_p \sim \mathcal{N}(0, R_p),
\end{align}
where $I_p$ is the scene irradiance, $F_p$ is the raw camera response, $CRF$ is the camera response function, $\mu_p$ is the noise in the scene irradiance and $R_p$ is the covariance of the noise.

\begin{figure}[!ht]
    \centering
    \newcommand{\imgfigWidth}{0.45\linewidth}
    \newcommand{\thisfigWidth}{0.45\linewidth}
    \begin{tabular}{M{\imgfigWidth}M{\imgfigWidth}}
        \includegraphics[trim={1cm, 1cm, 2cm, 1cm},clip,width=1.0\linewidth]{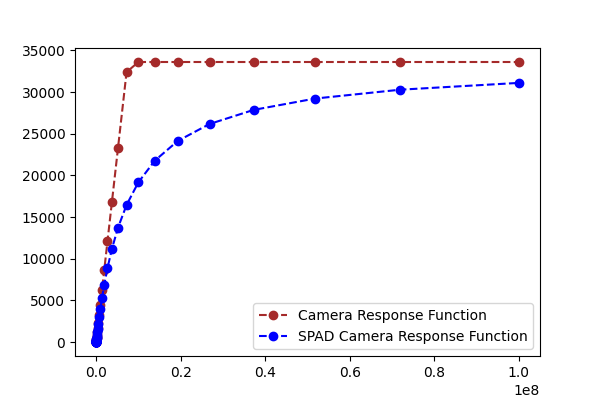} &
        \includegraphics[trim={1cm, 1cm, 2cm, 1cm},clip,width=1.0\linewidth]{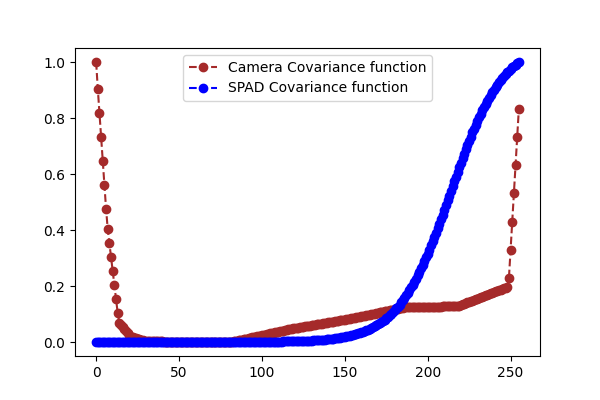}\\
        \footnotesize (a) Camera Response Function & \footnotesize (b) Covariance Function\\
    \end{tabular}
    \caption{ 
        Modelling the uncertainty of SPAD and standard camera (b) using the Camera Response Function (a) measured across different illumination on x-axis. 
        }
    \label{fig:crf}
\end{figure}
\noindent \textbf{SPAD Response Function}
The SPAD sensor response function is non-linear and is given by \Eq \ref{eq:spad}.
The noise in the SPAD sensor can be attributed to three main sources: (a) dark count noise (b) shot noise and (c) quantization noise.
The dark count noise is the noise due to the thermal electrons in the SPAD sensor.
The shot noise in SPADs is the variance in the detected number of photon count. 
This value monotonically increases with increasing brightness, reaches a maximum and then decreases at very high flux \cite{Ingle21CVPR}.
The difference between the noise characteristics of a conventional camera and SPAD is shown in \Fig \ref{fig:crf}.
It is clear from \Eq \ref{eq:spad} that in in low light, the camera response function of SPAD and conventional cameras are similar.
However, in bright illumination conditions, the SPAD sensor has a soft saturation, whereas the conventional camera has a hard saturation.
This is modelled in the covariance of the noise assicated to camera response ($\bar R$) in the SPAD sensor as shown in \Fig \ref{fig:crf}.
The covariance of noise $R(t)$ associated to the flux is given by:
\begin{equation}
    R (t) = \frac{\bar R}{\phi(t) + N_0}
    \label{eq:covupdate}
\end{equation}
where $\phi (t)$ is the flux calculated using \Eq \ref{eq:spad}.

\subsubsection{Asynchronous Kalman Filter}
In this section, we describe how to combine the uncertainty models of event camera and SPADs using Kalman filter.
Similar to \cite{wang21iccv}, we use a continuous time stochastic model of the log intensity.
\begin{align*}
    d N= e(t) dt + dw \\
    \hat{N}(t) =  N(t) + \mu
\end{align*}
where $dw$ is a Wiener process (continuous time stochastic process) and $\mu$ is the SPAD noise.
Solving this equation boils down to the ordinary differential equation as :
\begin{equation}
    \dot {N} (t) = e(t) - K (t) [\hat{N}(t) -{N}(t) ]
\end{equation}
where $K(t)$ is the Kalman gain defined below \ref{eq:kalmangain}.
When an event arrives, the filter state is updated as :
\begin{equation}
    \hat N(t) = \hat N(t-1) + e(t)
    \label{eq:ev_stateupdate}
\end{equation}

We compute per pixel gain $K(t)$ from state estimate and uncertainties as:
\begin{align}
    K(t) &= P(t) R(t)^{-1}
    \label{eq:kalmangain}
\end{align}
where $P(t)$ is the state covariance and $R(t)$ is the SPAD uncertainty covariance \cite{wang21iccv}.
At every new event, the state covariance $P(t)$ is updated from the previous timestamp $t-i$ as:
\begin{equation}
    P(t) = \frac{1}{P^{-1} (t-i)+ R^{-1}(t)(t-i)} + Q (t)
    \label{eq:stateupdate}
\end{equation}
where $Q$ is the event noise covariance \ref{eq:ev_noiseco}.
The AKF algorithm is summarized in \Alg \ref{algo:akf}.

\begin{algorithm}
    \caption{Event-SPAD Fusion Using Asynchronous Kalman Filter}
    \begin{algorithmic}[1]
    \State Initialise variables
    \For{New $i^{th}$ event at pixel $p$, $e(t_i)$}
        \If{new SPAD frame arrives}
            \State Deblur new SPAD frame based on \Eq \ref{eq:nedi} to obtain $\hat{N}(t_i)$
        \EndIf
        \State Update image covariance $R(t_i)$ using \Eq \ref{eq:covupdate} %
        \State Update state $\hat{N}(t_i)$  using \Eq \ref{eq:ev_stateupdate}
        \State Update covariance $P(t_i)$ using \Eq \ref{eq:stateupdate}
        \If{publishing new image}
            \For{all pixels $q$}
                \State Update state $\hat{N}(t_i^q)$
                \State Update covariance $P(t_i^q)$ 
                \State Write image
            \EndFor
        \EndIf
    \EndFor
    \end{algorithmic}
\end{algorithm}
\label{algo:akf}
\vspace{-2em}
\subsection{Adaptive SPAD sampling}
\label{sec:adaptive}
The SPAD sensor is a photon counting sensor that can detect single photons.
However this comes at the cost of high bandwidth and redundant data.
Instead, we propose an adaptive sampling mode for the SPAD sensor that captures a frame only when the uncertainty in state estimation is high.
The AKF approach described in \Sec \ref{sec:akf} not only produces a latent image but also provides an estimate of the uncertainty in the latent image.
We use this uncertainty to decide when to capture a new frame.
The uncertainty in the latent image is given by the covariance matrix $P(t)$.
We define the uncertainty $U$ in the latent image as the trace of the covariance matrix.
We set a threshold for the uncertainty and capture a new frame only when the uncertainty exceeds this threshold.
This adaptive sampling mode reduces the amount of redundant data captured by the SPAD sensor and also reduces the bandwidth required for data transmission.

\vspace{-1em}
\section{Experiments}
\label{sec:exp}
This section evaluates the performance of our proposed method on the task of image reconstruction from SPAD and event camera data.
We first introduce the simulation environment and hardware setup used for the synthetic and real-world experiments respectively.
Following which we describe the baseline methods used for comparison.
Then we perform experiments on synthetic data to show the effectiveness of our method in controlled settings to quantify the accuracy of our approach.
We then evaluate our method on real sensor data to show the practicality of our approach in real-world settings.
Finally, we perform an ablation study to show the effectiveness of each component of our method.
\begin{figure}
    \includegraphics[trim={0cm 14.2cm 0cm 0cm},clip,width=\linewidth]{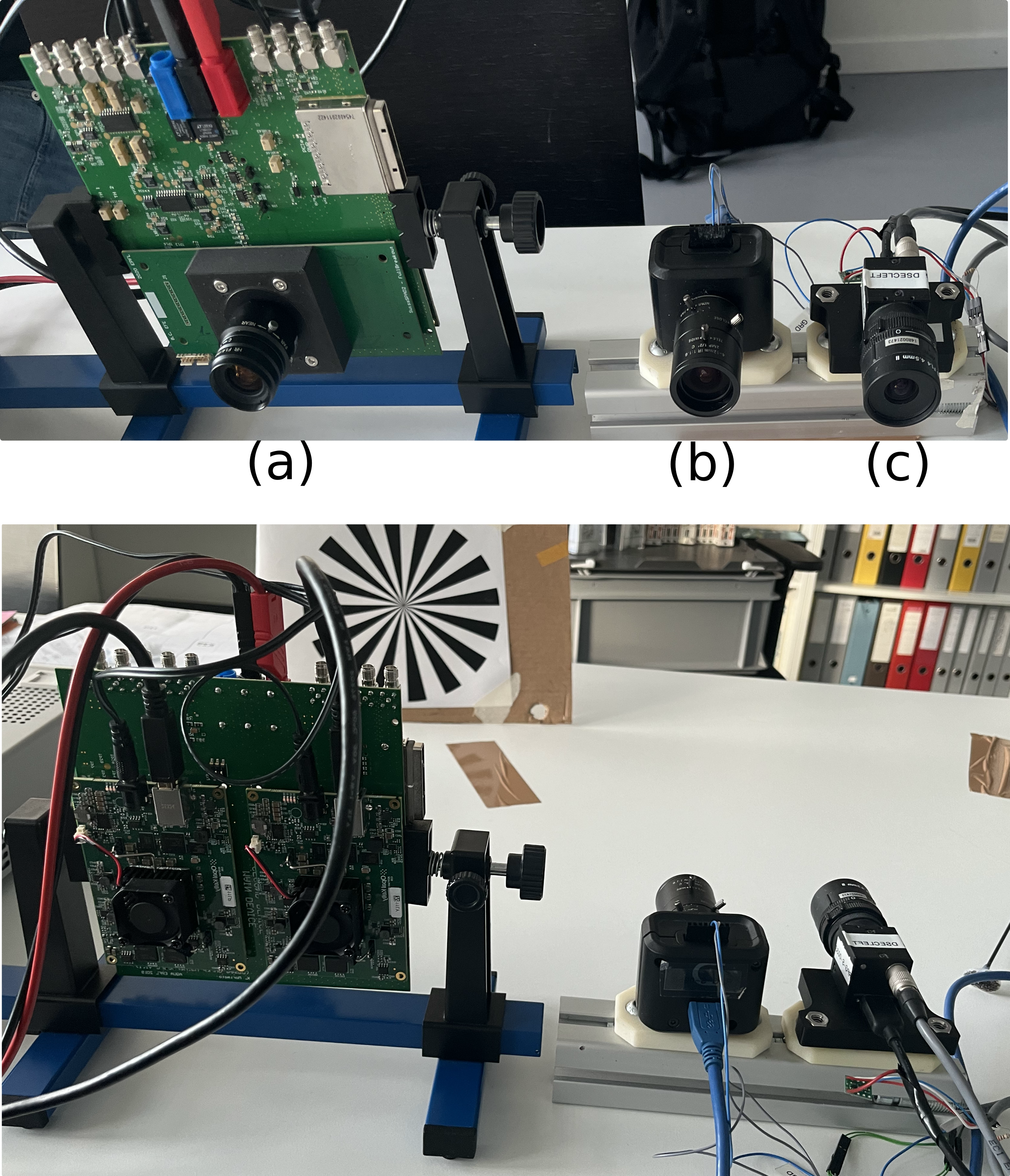}
    \vspace{-2em}
    \caption{\textbf{Experimental setup.} The top row shows the sensor setup consisting of (a) SwissSPAD2 \cite{ulku19jstqe}, (b) Prophesee event camera\cite{Finateu20isscc} and FLIR BlackFly S RGB camera from left to right.%
    }
    \vspace{-1em}
    \label{fig:setup}
\end{figure}
\vspace{-1em}
\subsection{\simdset Dataset}
\label{subsec:simulation}
We first evaluate our method on synthetic data to show the effectiveness of our method in controlled settings.
We use the SPAD simulation framework described in \cite{Ingle21CVPR} and event camera simulation framework described in \cite{Hu21cvprw} to generate synthetic data using the same parameters as the real sensor data.
The input to both the simulation frameworks is an HDR image and camera trajectory, which is used to render high framerate images to simulate the SPAD and event data.
The simulator also considers the noise characteristics of the SPAD and event camera sensors to generate the binary frames and events.
While the camera trajectory simulates the motion of the camera, we also provide illumination constraints in the forms of photons per sec to the simulator to simulate low light conditions.
The groundtruth intensity frames are also provided by the simulator, which are used for evaluation.

\vspace{-1em}
\subsection{\tldata Dataset}
\label{subsec:real}
We evaluated our method on real sensor data to show the practicality of our approach in real-world settings.
To the best of our knowledge, there is no publicly available dataset with synchronized SPAD and event camera data.
Therefore, we collect our own dataset using a state-of-the-art SPAD and event camera.
The experimental setup is shown in \Fig \ref{fig:setup}.
The SPAD camera\cite{ulku19jstqe} used has a resolution of $512 \times 512$ pixels, a maximum framerate of $\SI{100}{\kilo Hz}$ and exposure time of $\SI{10}{\micro \sec}$.
We use Prophesee Gen4 event camera\cite{Finateu20isscc} which has a resolution of $1280 \times 720$ pixels.
The two cameras were placed side by side and synchronized using a software trigger.
The alignment of the two cameras was challenging due to the fragility of the SPAD prototype sensor. 
For each sequence, the alignment was performed using feature-based homography estimation.
To achieve this, we first reconstructed images from events using \cite{Rebecq2021, Muglikar2021CVPR} and then estimated global homography by matching the points in the E2VID and SPAD image sequences.
A limitation of this calibration is that it can only align images upto a plane, which results in misalignment at the edges of the image in certain sequences. 
We only evaluate our method on the aligned regions of the images.

To compare the sensors with traditional camera, we used a FLIR Blackfly S global shutter camera with a resolution of $1080 \times 1440$ pixels. The exposure time of the FLIR camera was set to $\SI{10}{\milli \sec}$.
We collected over 5 sequences of different scenes with varying illumination conditions, camera motion and scene dynamics.
\vspace{-1em}
\subsection{Baseline Methods and Evaluation Metrics}
\label{subsec:baseline}
To the best of our knowledge, there are no existing methods that combine SPAD and event camera data for image reconstruction.
We therefore compare our method with the following baselines:
\begin{itemize}
    \item \textbf{CF (E)}: We use the event-only model of the complementary filter proposed in \cite{Scheerlinck18accv} to reconstruct continuous-time images from event data.

    \item \textbf{E2VID (E)}: We also use the state-of-the-art deep learning based event-based image reconstruction model proposed in \cite{Rebecq2021}.

    \item \textbf{Deblur (F)}: We use the state-of-the-art deblurring approach \cite{chu22cvprw} with traditional camera images.
    \item \textbf{AKF (F + E)}: We use the complementary filter proposed in \cite{wang21iccv} to combine the event data with the traditional camera data. %
    \item \rebuttal{\textbf{EvLight (F + E)}: We use the state-of-the-art event-guided low-light enhancement of images proposed in \cite{liang24cvpr}. The model is trained on SDE dataset \cite{liang24cvpr}.}
    
    \item \textbf{Deblur (S)} We use a state-of-the-art deep learning \cite{chu22cvprw} approach which was trained to denoise and deblur images. The motion blurred SPAD images are deblurred using this method.
    \item \textbf{QBP (S)}Quanta burst photography \cite{Ma20TOG} approach uses binary SPAD frames to align produce a deblurred image.   
    \item \rebuttal{\textbf{EvLight (S + E)}: We repurpose the event-guided low-light enhancement of images \cite{liang24cvpr} to use the SPAD images instead of traditional camera images. The model is trained on SDE dataset \cite{liang24cvpr}.}
\end{itemize}
Note, that next to each method, we also mention the sensor data used as input, where S denotes SPAD, E denotes event camera and F denotes traditional camera and the combination is denoted by the `+' sign.
To evaluate all the methods, we use the traditional image reconstruction metrics PSNR (peak signal to noise ratio), which captures the effect of noise in the reconstructed image. 

In addition to these metrics, we also evaluate the performance of the sensors using modulation transfer function (MTF) metrics on the real sensor data.
MTF is a measure of the ability of an imaging system to faithfully transfer spatial frequencies from the object to the image.
The setup consists of a siemens star target (as shown in \Fig \ref{fig:setup}), which is imaged by the all the three sensors namely, SPAD, event camera and traditional camera as the star is rotated along the center.
The siemens star target has a series of concentric rings with increasing spatial frequency going from the edge to the center.
In an ideal sensor, the MTF should be close to 1 for all spatial frequencies.
However, in practice, the MTF decreases with increasing spatial frequency due to the finite size of the pixels and the motion blur.
We use the MTF metrics to evaluate the performance of the sensors in capturing high frequency details in the presence of motion blur.

\global\long\def\figWidth{0.3\linewidth}
\begin{figure}[!h]
	\centering
    \setlength{\tabcolsep}{2pt}
	\begin{tabularx}{\linewidth}{
	M{\figWidth}
	M{\figWidth}
	M{\figWidth}}
        \includegraphics[width=\linewidth]{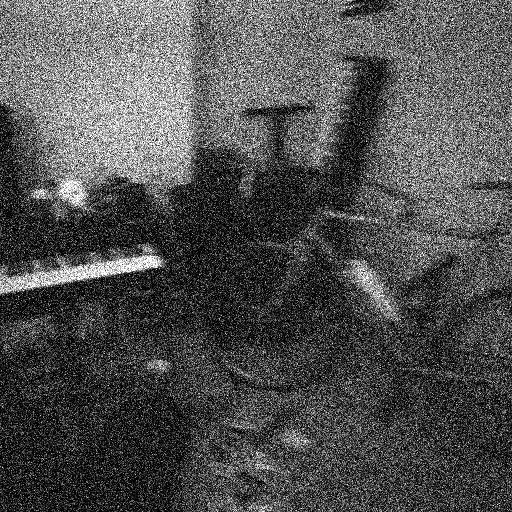} &
        \includegraphics[width=\linewidth]{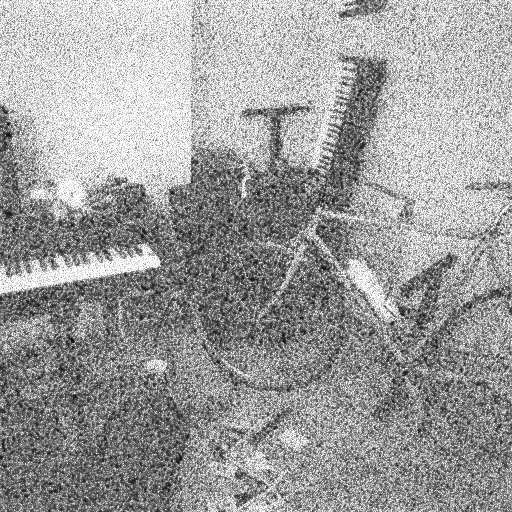} &
        \includegraphics[width=\linewidth]{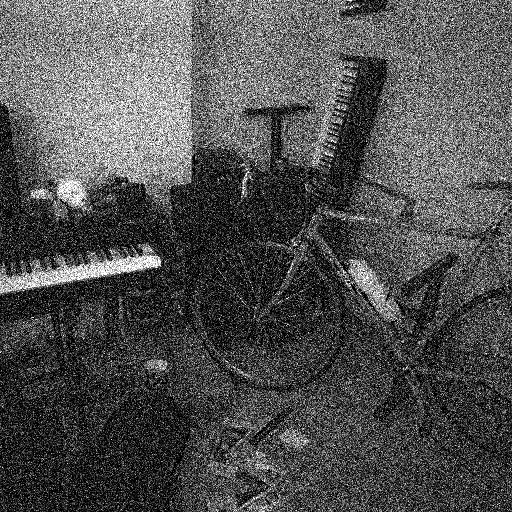} \\
        \footnotesize (a) Frame \footnotesize & (b) Frames + Events \cite{liang24cvpr} &  \footnotesize (c) Frames + Events \cite{pan19cvpr}\\
        \includegraphics[width=\linewidth]{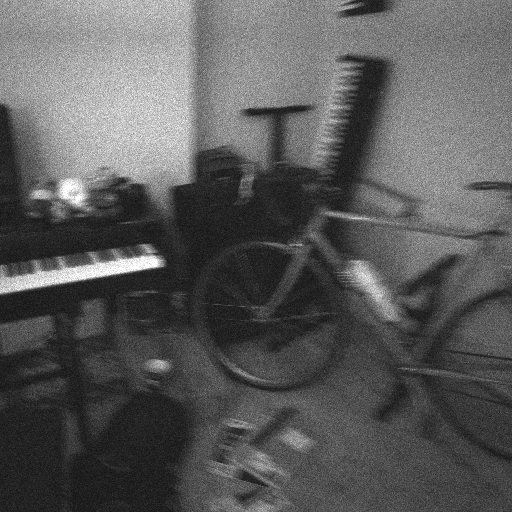} &
        \includegraphics[width=\linewidth]{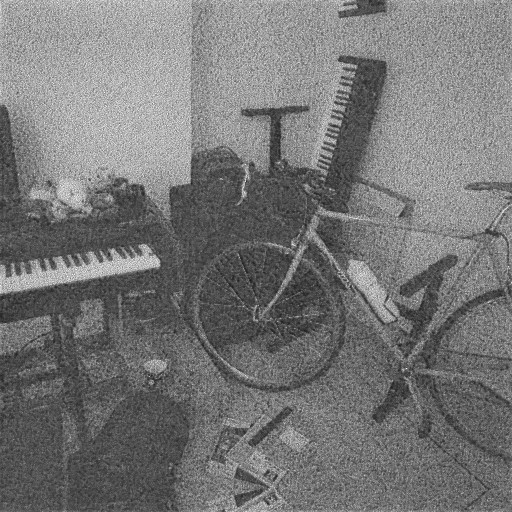} &
        \includegraphics[width=\linewidth]{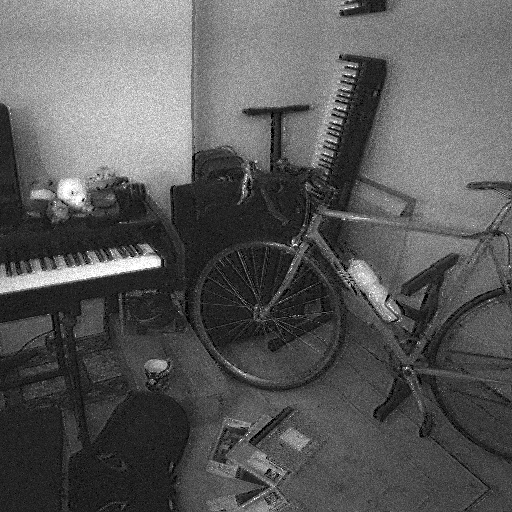}\\
        \footnotesize (d) SPAD &  \footnotesize (e) SPAD + Events (\cite{liang24cvpr}) &  \footnotesize (f) SPAD + Events (\textbf{Ours})\\
        \includegraphics[width=\linewidth]{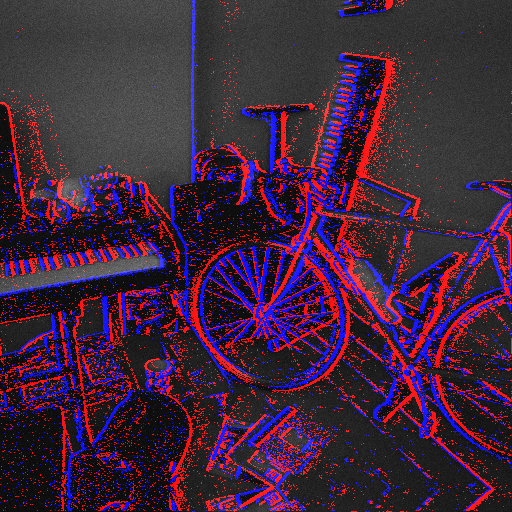}&
        \includegraphics[width=\linewidth]{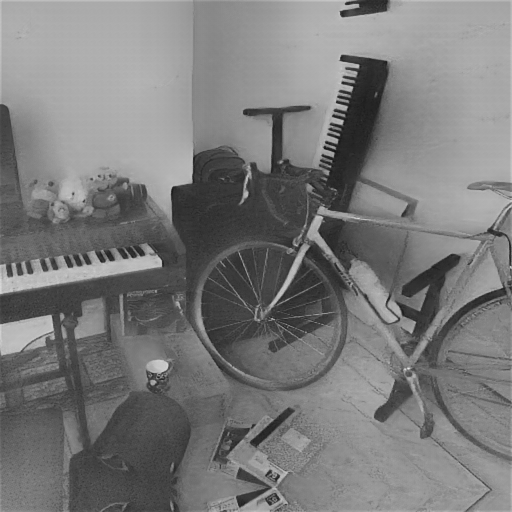}&
        \includegraphics[width=\linewidth]{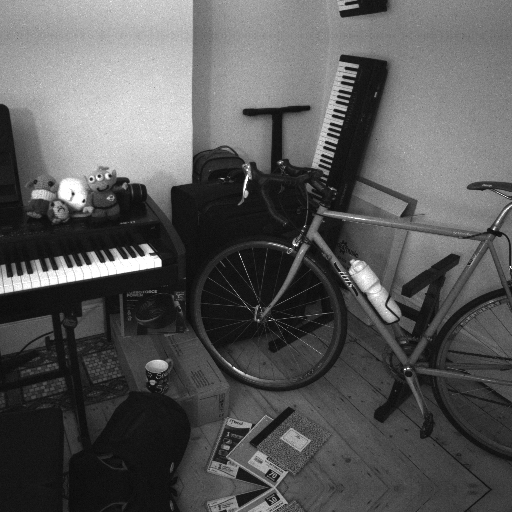}\\
        \footnotesize (g) Events &  \footnotesize (h) Event-Image (\cite{Rebecq2021}) &  \footnotesize (i) Groundtruth\\
    \end{tabularx}
    \caption{\textbf{Simulated Results.} 
    Comparing traditional camera (a), SPAD camera (d), and event camera (g) in low-light scenes.
    Our method (f) combines events with SPADs resulting high quality images with low noise compared to frames+event baseline.
    }
    \label{fig:sim:frame_spad_event}
\end{figure}
\vspace{-2em}
\section{Results}
\label{sec:results}

\subsection{\simdset Dataset}
\label{subsec:synthetic}
We first evaluate our method on synthetic data to show the effectiveness of our method in controlled settings and the metrics are reported in \Tab \ref{tab:sim:results}.
The naive integration baseline results in significant motion blur in the reconstructed image.
In contrast, our method is able to reconstruct the image with sharper edges as can be seen by the keyboard keys (row $1$) and carpet patterns (row $2$).
Additionally, our approach overall has lower noise compared to individual binary frames.

\begin{table}[h!]
    \centering
    \begin{adjustbox}{max width=\linewidth}
    \begin{tabular}{@{}llllll@{}}
    \toprule
        & &  & \textbf{Office} & \textbf{Piano} & \textbf{Yucca} \\
        Sensor & Method &  Bandw. $\downarrow$& PSNR (dB)$\uparrow$ & PSNR (dB)$\uparrow$ & PSNR (dB) $\uparrow$ \\ \midrule
        E & $CF_e$\cite{Scheerlinck18accv} & & 10.48           & 9.45           & 11.39          \\
        & $E2VID$\cite{Rebecq2021}          & & 13.50           & 12.62           &  15.43        \\ \midrule
        F & Deblur \cite{chu22cvprw}         & 0.2 & 5.62           & 5.25           & 5.86           \\ %
        F + E & $AKF$\cite{wang21iccv}       &  0.2 & 14.15           & 13.9           & 13.67          \\ 
        \rebuttal{F +E} & \rebuttal{EvLight} \cite{liang24cvpr} & \rebuttal{0.2} & \rebuttal{9.71}  & \rebuttal{8.93} & \rebuttal{9.7 } \\ \midrule
        S & Deblur \cite{chu22cvprw}          & 0.4 & 23.17           & 19.99           & 21.86          \\
        S & QBP SPAD \cite{Ma20TOG}           & 100 &  11.23           & 17.38           & 18.48          \\% \midrule
        \rebuttal{S +E} & \rebuttal{EvLight} \cite{liang24cvpr} & \rebuttal{0.4} & \rebuttal{9.84} & \rebuttal{9.1} & \rebuttal{9.66} \\ \midrule
        S + E & Ours               & 0.4 & \textbf{21.50}     & \textbf{19.44}    & \textbf{21.28}    \\ 
        S + E & Ours (Adapt.)     & \textbf{~0.1} & 17.45     & 18.04    & 17.8    \\ \bottomrule
    \end{tabular}
    \end{adjustbox}
    \medskip
    \caption{PSNR (dB) and Bandwidth (kHz/pixel) across different sensors on the \simdset dataset. The best performance is highlighted in bold.}
    \label{tab:sim:results}
    \vspace{-2em}
\end{table}

\noindent \textbf{Comparison of SPAD to Traditional Camera}
In low light scenarios, the SPAD sensor is able to capture the scene with higher dynamic range compared to the traditional camera.
This can be seen in \Fig \ref{fig:sim:frame_spad_event} (a) and (b), where the output of the SPAD sensor is less noisy compared to frame camera for the same exposure time of $\SI{10}{\micro \sec}$.
Given the long exposure time, both these sensors suffer from motion blur, which further degrade the quality of the captured scene.
In contrast, the event camera does not suffer from motion blur, as shown in \Fig \ref{fig:sim:frame_spad_event} (c).
The combination of event camera with frames in low light scenarios results in high quality images, as shown in \Fig \ref{fig:sim:frame_spad_event} (d), however is still noisy due to inherent noise of the frame sensor.
Combining event camera with SPADs on the other hand results in high quality images with low noise, as shown in \Fig \ref{fig:sim:frame_spad_event} (e) when comparing to the groundtruth image in \Fig \ref{fig:sim:frame_spad_event} (f).
We also show quantitative results in \Tab \ref{tab:sim:results} to compare the performance of combining event camera with SPADs to combining events with traditional cameras in low light.
\rebuttal{We also evaluate the performance of state-of-the-art event-guided low-light image reconstruction approach proposed in \cite{liang24cvpr} on our dataset.
Qualitative and quantitative results show  our method outperforms the event-guided approach in all scenes.
Although the event-guided approach is able to recover some of the details in the scene, it still struggles due to the inherent noise of the frame sensor.
Using the higher-quality images coming from SPADs, the same method gives higher quality results. The qualitative results are shown in \Fig \ref{fig:sim:frame_spad_event}.}
We can achieve upto $7dB$ improvement in PSNR indicating that SPADs provide better image quality compared to traditional cameras in low light scenarios.

\begin{figure}[h]
\centering
\colorlet{crystal}{blue!75}
\colorlet{red}{red!50}
\colorlet{lightgreen}{green!75}
\pgfplotstableread{
1	10.45   10.452   14.47  
2	11.39   13.67   21.28   
3	9.7   18.78   19.22  
}\dataset
\tikzstyle{every node}=[font=\footnotesize]
\begin{tikzpicture}
\begin{axis}[ybar,
        width=8.3cm,
        height = 4.9cm,
        bar width = 3mm,
        ymin=0,
        ylabel shift = -4pt,
        ylabel={PSNR (dB)},
        xtick=data,
        xticklabels = {
            100 Lux ,
            1000 Lux,
            10000 Lux,
        },
        ymajorgrids,
        major x tick style = {opacity=0},
        minor x tick num = 0,
        minor tick length=-1ex,
         enlarge x limits={abs=0.3},
         legend columns = 3,
         legend style = {anchor=south east}, 
         legend image code/.code={
        \draw [#1,draw=none] (0cm,-0.1cm) rectangle (0.2cm,0.25cm); },
        ]
\addplot[draw=none,fill=red] table[x index=0,y index=2] \dataset; %
\addplot[draw=none,fill=lightgreen!50] table[x index=0,y index=3] \dataset; %
\legend{AKF \cite{wang21iccv} (F+E), Ours (S+E)}%
\end{axis}
\end{tikzpicture}
\caption{\textbf{Effect of illumination on sensor fusion.} Using SPAD output improves robustness of AKF approach \cite{wang21iccv} in low light.}
\label{fig:sim:illumination}
\vspace{-2em}
\end{figure}
\noindent \textbf{Effect of Illumination}
We compare the effect of illumination on the sensor fusion capabilities of SPADs, frames and events.
We evaluate the performance of our method and frames and event fusion \cite{wang21iccv} on \simdset while varying illumination conditions starting from $\SI{100}{\lux}$ to $\SI{10000}{\lux}$.
The quantitative results are shown in \Fig \ref{fig:sim:illumination}.
At low illumination, the SPAD sensor is able to capture the scene with higher dynamic range compared to the traditional camera.
Therefore, SPAD-based approaches tend to outperform traditional camera-based approaches in low light scenarios.
We show that the combination of SPADs and event cameras results in highest image reconstruction accuracy compared to frame-based counterpart at all illumination levels.
However, as the illumination increases the performance between the SPADs and frame counterpart decreases as both sensors tend to saturate

\subsection{\tldata Dataset}
\label{subsec:realresults}
We now evaluate our method on real sensor data collected using the setup described in \Sec \ref{subsec:real}.
We first provide quantitative and qualitative evaluation on the effect of motion blur on SPADs using MTF analysis in \Sec \ref{sec:exp:mtf}.
Following this we show results on natural scenes captured by the sensors in \Sec \ref{sec:exp:spadwild}

\noindent \textbf{MTF Analysis}
\label{sec:exp:mtf}
The modulation transfer function (MTF) evaluates the ability of an imaging system to faithfully transfer spatial frequencies from the object to the image.
It is a measure of the sharpness of the image and is used to evaluate the performance of the sensors in the presence of scene motion.
With a static scene, the MTF of a SPAD camera remains constant around $0.55$ across different spatial frequencies as shown in \Fig \ref{fig:real:mtf}.
However, introducing scene motion induces motion blur decreasing the MTF of the SPAD down to $0.37$.
Deblurring the SPAD images using \cite{chu22cvprw} improves the MTF slightly, however the best performance is achieved by our method with MTF of $0.51$ getting very close to the MTF of the static scene.
Qualitative results of the MTF analysis are shown in \Fig \ref{fig:real:mtfqual}.

\global\long\def\figWidth{0.45\linewidth}
\begin{figure}[!ht]
	\centering
    \setlength{\tabcolsep}{2pt}
	\begin{tabularx}{\linewidth}{
        M{\figWidth}
        M{\figWidth}}
        \footnotesize (a) NI (S)  &\footnotesize (b) Deblur(S) \cite{chu22cvprw} \\
        \includegraphics[trim={0cm, 5cm, 10cm, 5cm},clip,width=1.0\linewidth]{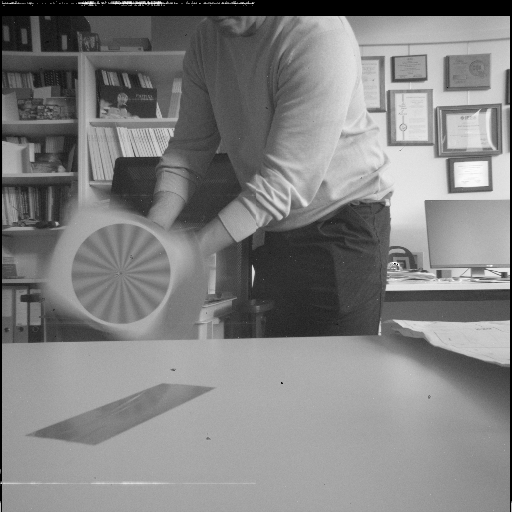}&
        \includegraphics[trim={0cm, 5cm, 10cm, 5cm},clip,width=1.0\linewidth]{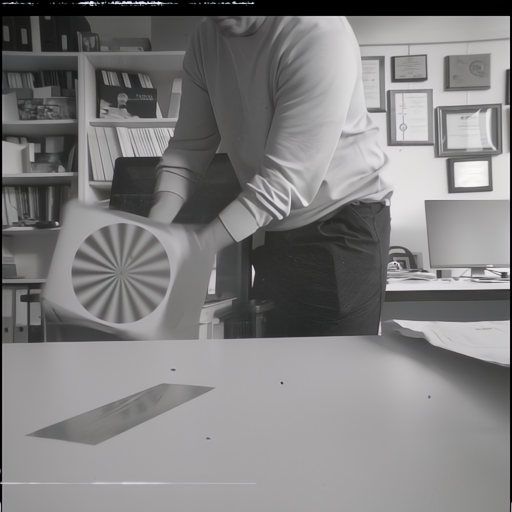} \\
        \footnotesize  (c) Ours(S+E) & \footnotesize (d) GT (S) \\
        \includegraphics[trim={0cm, 5cm, 10cm, 5cm},clip,width=1.0\linewidth]{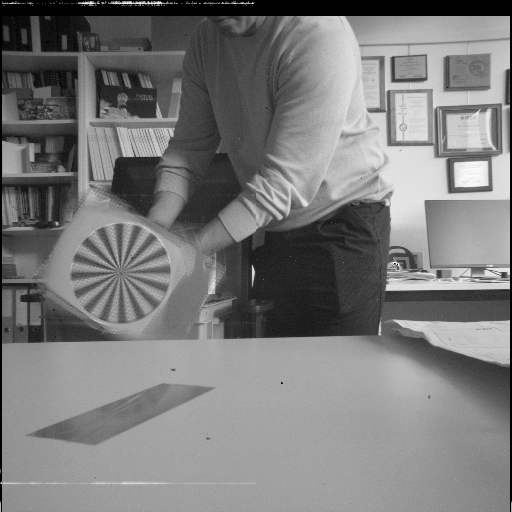} & 
        \includegraphics[trim={0cm, 5cm, 10cm, 5cm},clip,width=1.0\linewidth]{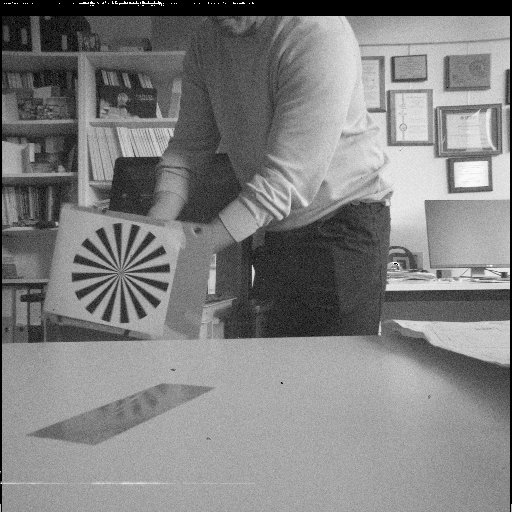}\\
        \\
    \end{tabularx}
    \caption{ \textbf{Effect of scene motion on image quality}
        Using MTF analysis on Siemens star, we observe our method (c) preserves the contrast edges and texture details better than SPAD-only methods NI(a) and  Deblur\cite{chu22cvprw} (b). For comparison, (d) shows the groundtruth static Siemens star capture by SPAD. 
        }
    \label{fig:real:mtfqual}
\end{figure}

\begin{figure}[h]
\centering
\colorlet{crystal}{blue!75}
\colorlet{red}{red!50}
\colorlet{lightgreen}{green!75}
\pgfplotstableread{

2 0.54  0.42  0.0 0.44  0.52
3 0.55  0.42  0.0 0.43  0.51
4 0.54  0.42  0.0 0.41  0.51
5 0.50  0.41  0.0 0.39  0.49
6 0.50  0.38  0.0 0.37  0.49
}\dataset
\tikzstyle{every node}=[font=\footnotesize]
\begin{tikzpicture}
\begin{axis}[ybar,
        width=8.3cm,
        height = 4.9cm,
        bar width = 2mm,
        ymin=0,
        ylabel shift = -5pt,
        ylabel={MTF},
        xtick=data,
        xticklabels = {
            0.1,
            0.2,
            0.3,
            0.4,
            0.5,
        },
        ymajorgrids,
        major x tick style = {opacity=0},
        minor x tick num = 0,
        minor tick length=-1ex,
         enlarge x limits={abs=0.5},
         legend columns = 4,
         legend style = {anchor=south east}, 
         legend image code/.code={
        \draw [#1,draw=none] (0cm,-0.1cm) rectangle (0.2cm,0.25cm); },
        ]
\addplot[draw=none,fill=crystal!30] table[x index=0,y index=1] \dataset; %
\addplot[draw=none,fill=red] table[x index=0,y index=2] \dataset; %
\addplot[draw=none,fill=crystal!70] table[x index=0,y index=4] \dataset; %
\addplot[draw=none,fill=lightgreen!50] table[x index=0,y index=5] \dataset; %
\legend{Static (S), NI(S),  DeBlur (S)\cite{chu22cvprw}, Ours (S+E)}
\end{axis}
\end{tikzpicture}
\caption{MTF analysis of SPADs and event cameras across different spatial frequencies in lines per mm. Higher is better.
}

\label{fig:real:mtf}
\end{figure}
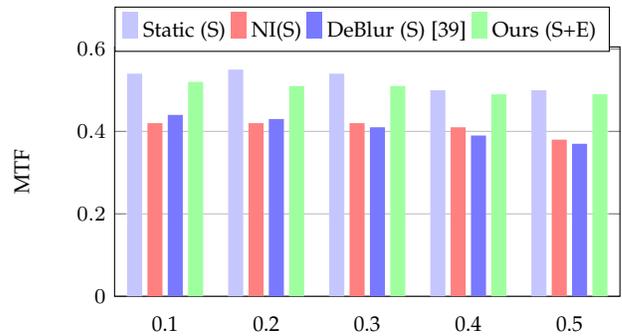
\noindent \textbf{SPAD in the wild}
\label{sec:exp:spadwild}
The qualitative results of our method on real sensor data are shown in \Fig \ref{fig:real:results}.
The naive integration baseline results in significant motion blur in the reconstructed image.
In contrast, our method is able to reconstruct the image with sharper edges as can be seen by edges of the Siemens star (row $1$) and the fan blades (row $3$).
Moreover, our approach overall has lower noise compared to individual binary frames seen in the tunnel sequence.

\global\long\def\figWidth{0.18\linewidth}
\begin{figure}[!t]
	\centering
    \setlength{\tabcolsep}{2pt}
	\begin{tabularx}{\linewidth}{
        M{0.2cm}
        M{\figWidth}
        M{\figWidth}
        M{\figWidth}
        M{\figWidth}
        M{\figWidth}
        }
        & \footnotesize (a) Events & \footnotesize(b) Deblur(S) \cite{chu22cvprw}& \footnotesize(b) E2VID(E) \cite{Rebecq2021} & \footnotesize(c) Ours (S+E) & \footnotesize (d) QBP\\

        \rotatebox{90}{\makecell{Fan}}&
        \includegraphics[trim={3cm, 5cm, 10cm, 6cm},clip, width=1.0\linewidth]{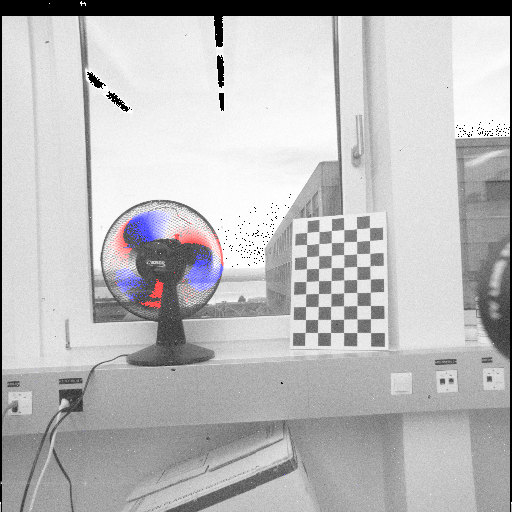} &
        \includegraphics[trim={3cm, 5cm, 10cm, 6cm},clip,width=1.0\linewidth]{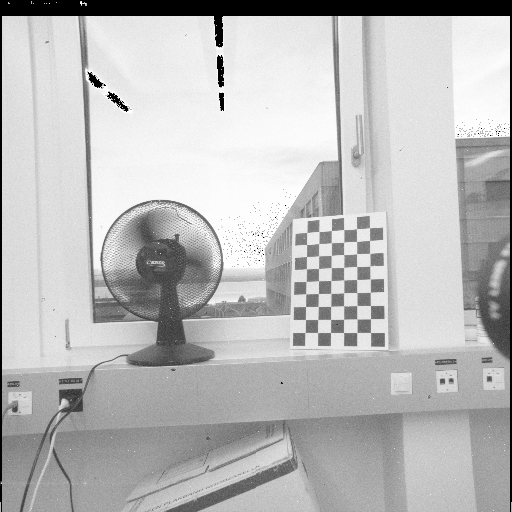} &
        \includegraphics[trim={3cm, 5cm, 10cm, 6cm},clip,width=1.0\linewidth]{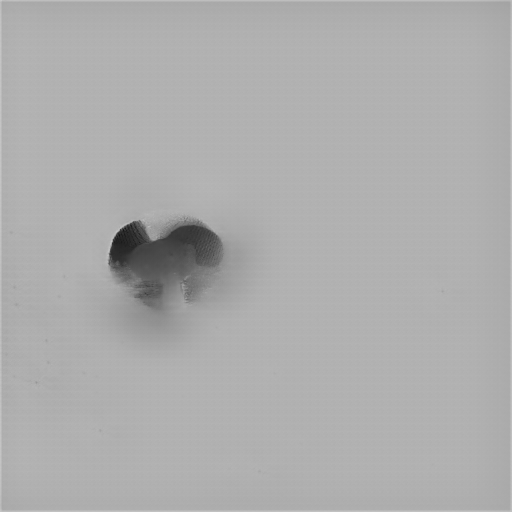} &
        \includegraphics[trim={3cm, 5cm, 10cm, 6cm},clip, width=1.0\linewidth]{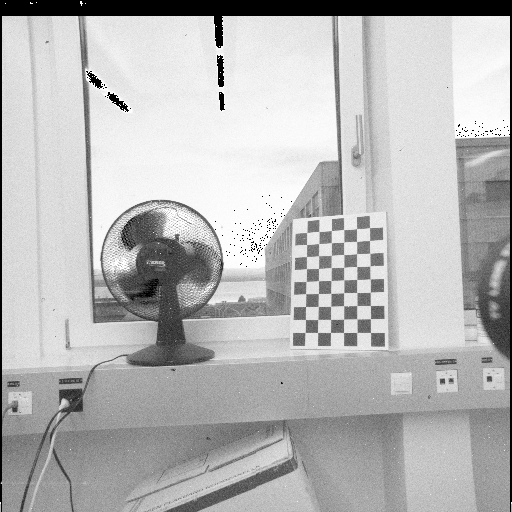} &
        \includegraphics[trim={3cm, 5cm, 10cm, 6cm},clip, width=1.0\linewidth]{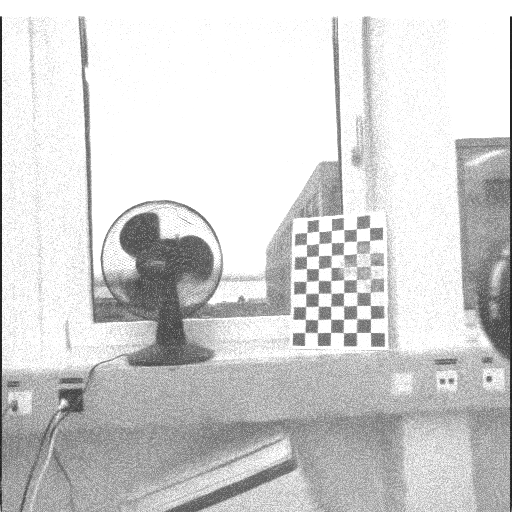}
        \\

        \rotatebox{90}{\makecell{Zoom}}&
        \includegraphics[trim={4.5cm, 9cm, 12cm, 7cm},clip, width=1.0\linewidth]{images/fig5/fan/1/event_00003.png.png} &
        \includegraphics[trim={4.5cm, 9cm, 12cm, 7cm},clip,width=1.0\linewidth]{images/fig5/fan/1/spad_2_00004.png} &
        \includegraphics[trim={4.5cm, 9cm, 12cm, 7cm},clip,width=1.0\linewidth]{images/fig5/fan/1/e2calib_0000000000311900000.png} &
        \includegraphics[trim={4.5cm, 9cm, 12cm, 7cm},clip, width=1.0\linewidth]{images/fig5/fan/1/ours_00011.png} &
        \includegraphics[trim={4.3cm, 9cm, 12.2cm, 7cm},clip, width=1.0\linewidth]{images/fig5/fan/1/qbp.png} 
        \\

    \end{tabularx}
    \caption{ Qualitative results comparing the best SPAD-only baseline \cite{chu22cvprw} (b), event-only baseline $E2VID$ \cite{Rebecq2021} (c) and our method (c) on \tldata. The aligned and synchronized events are overlaid on SPADs images and visualized in (a). 
        }
    \label{fig:real:results}
    \vspace{-1em}
\end{figure}

\noindent \textbf{Comparison to Quanta Burst Photography (QBP)}
\label{sec:exp:qbp}
\begin{table}[h!]
    \centering
    \begin{tabular}{@{}llllll@{}}
    \toprule
    \textbf{Method} &\textbf{Sensor} & \multicolumn{4}{c}{\textbf{Bandwidth}} \\  \midrule
     & & \textbf{1/5} & \textbf{1/10} & \textbf{1/20} & \textbf{1/30} \\  \midrule
    QBP & SPAD & 19.402 & 19.900 & 17.341 & 15.296 \\ 
    Ours & SPAD + Events &\textbf{24.806} & \textbf{22.619} & \textbf{18.916} & \textbf{16.991} \\ \bottomrule
    \end{tabular}
    \medskip
    \caption{PSNR values for QBP and AKF methods across various bandwidths.}
    \label{tab:qbp_vs_akf}
    \vspace{-2em}
\end{table}
We compare our method to the QBP \cite{Ma20TOG} on the fan sequence of \tldata dataset.
We evaluate the effect of reducing the bandwidth of the SPAD sensor on both methods.
Reducing the bandwidth, implies fewer SPAD images used for both QBP and our method resulting noisier reconstruction.
However, since QBP requires all the binary frames to align the images, the increased noise has a significant impact on the performance.
In our case, since we rely on events for low bandwidth deblurring, the noise has a lesser impact on the performance.
The quantitative results are shown in \Tab \ref{tab:qbp_vs_akf}.
Detailed analysis is provided in \supp

\vspace{-1em}
\section{Discussion}
\label{subsec:discussion}
We have introduced a novel sensor fusion approach for capturing high-speed, HDR scenes with low bandwidth.
We show that for the wider range of dynamic scenes, the sensor capabilities of SPADs and event cameras are complementary to each other.
Our approach leverages the high temporal resolution of event cameras and the HDR property of SPADs.
We demonstrated that the combination of SPADs and event cameras 
can achieve significant improvement compared to frame-based counterparts for low-light imaging.
Moreover, by using a Kalman filter approach for event-SPAD fusion, we can reduce the bandwidth of SPADs by up to $4$ times while outperforming the conventional camera baselines by $5$ dB.

\section*{Acknowledgment}
We thank Paul Mos and Dr.David Rodriguez Martinez for their help with the SPAD sensor. 
We also thank Nico Messikommer and Leonard Bauersfeld of the Robotics and Perception Group for their valuable feedback and discussions.%

\ifCLASSOPTIONcaptionsoff
  \newpage
\fi

\bibliographystyle{IEEEtran}
\bibliography{all}
\balance

\newpage



\section*{Supplementary material}

# Supplementary Material

## 1 Event SNR calculation

Due to the unique asynchronous and sparse nature of event camera, quantifying it's performance under different illumination and scene motion is challenging. SNR in the context of event camera depends not only on the illumination but also on the illumination change and the scene motion. Moreover, the noise of event camera is quite complex to model (and an active research area [1], [2]). This makes it difficult to compare the performance of event camera with traditional cameras and SPADs. We therefore simplify the problem to compare the performance of event camera with traditional cameras and SPADs in terms of SNR. Consider the scenario where the illumination is given by $\Phi$ and the illumination change is given by $\Delta\Phi$. The event camera parameters are given by the event trigger probability $P_e(\Phi, \Delta\Phi)$, the contrast threshold $C$ and the noise $N(\phi)$. The noise here corresponds to shot noise which is a function of illumination (i.e high at lower illumination and low at higher illumination) without any scene motion. The event trigger probability is a function that depends on both the illumination and the illumination change, encompassing the sensor noise artifacts. Thus, the SNR for event camera is given by:

$$\text{SNR}_{\text{events}}(\Phi, \Delta\Phi) = 10 \log_{10} \frac{\left(P_e * \frac{\Delta\Phi}{C}\right)}{N(\phi)} \quad (1)$$

Therefore, given event probability, contrast threshold and noise, we can compare the SNR of event camera with traditional cameras and SPADs. Estimating these values is challenging and is an open problem in the field of event camera research. In this paper, we resort to empirical measured values of event probability and noise. It was shown in [3], for fixed contrast threshold, increases illumination decreases the event probability and for a fixed illumination, the event probability increases with illumination change resulting in S-curves. This was measured for $4$ illumination levels of $2Lux$, $10Lux$, $100Lux$ and $600Lux$ and illumination changes of $5\%$ to $100\%$. We therefore use this empirical data to calculate the event trigger probability for different illumination and illumination change. Estimating the static noise as a function of illumination, we use the empirical data from [2]. In the absence of any motion, the noise as a function of illumination is an exponentially decreasing function. The noise here is measured as the event rate per pixel. This is intuitive as the noise is high at lower illumination and low at higher illumination. We can now combine these two empirical measurements to calculate the SNR of event camera for different illumination and illumination change. In Fig. 1, we compare the SNR of all 3 sensors: frame, SPAD

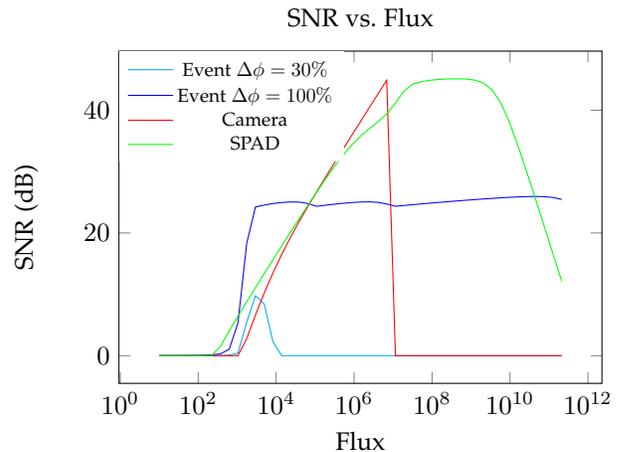

Fig. 1. SNR comparison between SPAD, event and frame sensors at different flux regimes. SPADs and Events have a better quality information at low lights compared to conventional cameras.

and event camera with increasing illumination. Detailed observations can be made from this plot as follows:

- Comparison between SPADs and frames: SPADs have a better low light performance [4]. At higher illuminations, the SNR of frames drops down due to the saturation of the pixels. SPADs on the other hand, have a better SNR at higher illumination levels because of their high dynamic range.
- Effect of illumination change on event SNR: Now only comparing the two event curves, one can see that higher contrast changes result in better SNR across illumination levels. This is intuitive as the event camera is able to capture more information about the scene with higher contrast changes. Smaller contrast changes can be captured at lower illuminations, but as the illumination increases, the SNR of the event camera drops down because of the logarithmic nature of event pixel.
- Comparing the three sensors, we can see that SPADs and events have a better SNR at lower illumination (with SPADs being more sensitive at lower illumination). At higher illumination, frames saturate resulting in a drop in SNR. SPADs on the otherhand, has a smoother drop in SNR at higher illumination. Events (for $100\%$ contrast change) do not have a drop in SNR even at high illumination.

## 2 Effect of Exposure

**SimSPAD** We also evaluate the performance of our method on synthetic data with varying integration time of SPADs.

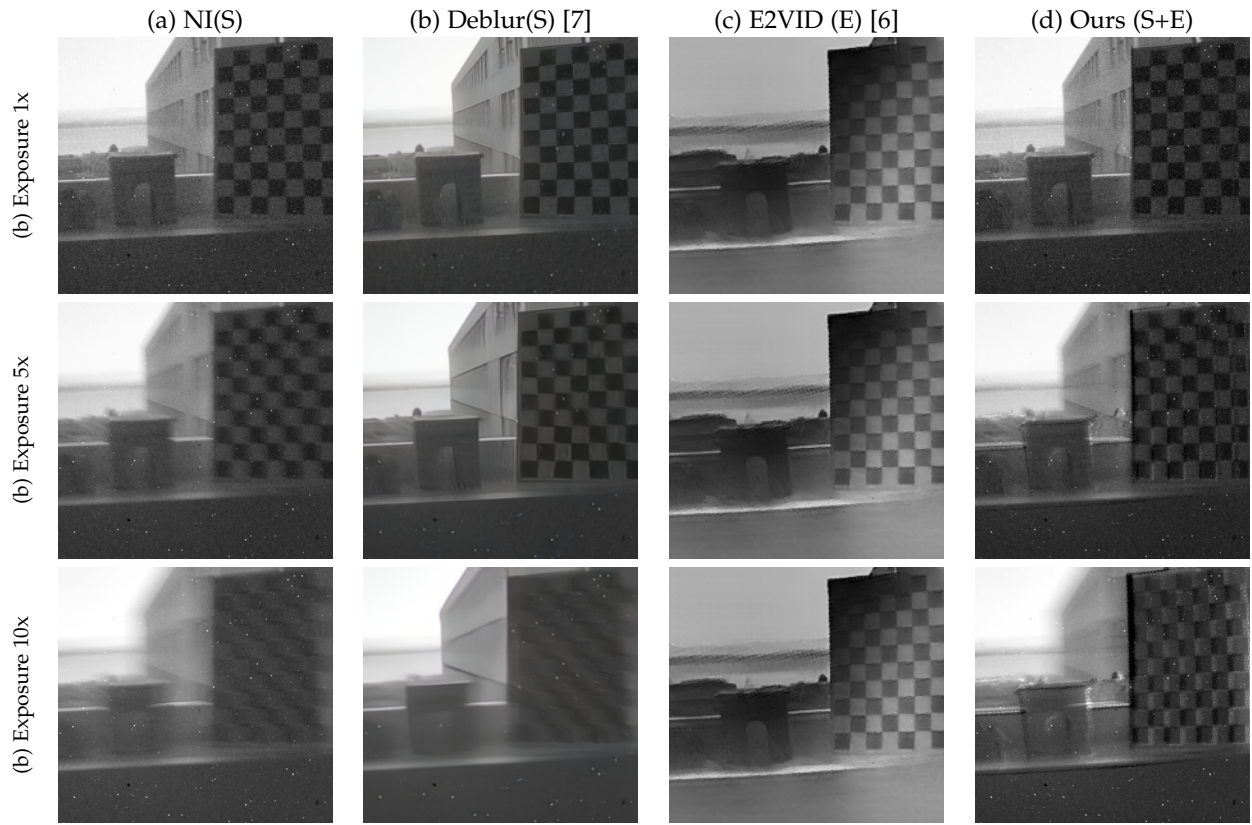

Fig. 2. Results on deblurring SPAD images at different exposure times using SPAD-only methods: NI (a), and Deblur [7] (b) and event-only baseline [6] (c) and ours (d).

Table 1 shows the performance of our method with varying exposure times. While increasing the integration time allows for lower bandwidth and higher SNR, it also results in more motion blur. At extremely low exposure times ($1\times$ and $5\times$), the noise is quite significant and the images do not have any meaningful information and therefore the deblurring approach results in a very low error for both SPADs and frames. At higher integration period of $10\times$, the SNR of SPAD is significantly better than frame-based method for the deblurring approach. This is exactly the non-linear behaviour of SPADs which leverage the low-light sensitivity better than frames. While the SNR is higher, it only reflects the noise quality. The SPAD image however is still significantly motion blurred. With our method, we outperform frame-based baseline by 8 dB for the highest exposure time.

| Sensor | Method | **1x** | **5x** | **10x** |
|---|---|---|---|---|
| Event | $CF_e$ [5] | 11.39 | 11.39 | 11.39 |
|  | $E2VID$ [6] | 15.43 | 15.43 | 15.43 |
| Frames | Deblur [7] | 5.86 | 5.86 | 5.86 |
| Frames + Events | $AKF$ [8] | 13.93 | 13.39 | 12.71 |
| SPAD | QBP SPAD [4] | 17.15 | 11.93 | 11.23 |
| SPAD | Deblur [7] | 5.87 | 5.57 | 23.17 |
| SPAD + Events | Ours | **15.23** | **18.48** | **21.5** |

TABLE 1
Effect of exposure time on simulated data: PSNR (dB) for increasing exposure time by factor of 1x, 5x and 10x.

**HS-ESPAD** Exposure time of SPADs have significantly impact the motion blur. We show that increasing the exposure time results in more motion blur in SPAD only methods, which our approach is resilient to. While the deblur method [7] is able to reduce motion blur of the window frame and the building facade, it is not able to reduce motion blur of the checkered board. On the otherhand, event-only method such as $E2VID$ [6], reconstructs the image reliably only where the contrast edges are strong. For example, the upper left corners of the checkerboard are not reliably reconstructed by $E2VID$.

| Sensor | Method | **1x** | **5x** | **10x** |
|---|---|---|---|---|
| Event | $E2VID$ [6] | 9.5 | 9.50 | 9.40 |
| SPAD | Deblur [7] | 15.73 | 15.87 | 18.57 |
| SPAD + Events | Ours | **25.7** | **21.16** | **20.25** |

TABLE 2
Effect of exposure time on real data: PSNR (dB) for increasing exposure time.

Our method on the other hand is able to reduce motion blur in all the sequences. Note that we can only deblur areas which overlap with the event camera field of view. Since the baseline between event camera and SPADs is different, not all the areas are deblurred resulting in artifacts in the background. We also show the quantitative results in Table 2. We also show qualitative results in Fig. 2. As events do not have exposure time, the exposure time has no effect on the performance of event-only methods. On the

other hand, increasing exposure time increases the motion blur, therefore the performance of SPAD only method decreases significantly at higher exposure times. Our method outperforms the SPAD only baseline by 1.68 dB at the $10\times$ exposure and 10 dB at the lowest exposure.

## 3 EVENT DOUBLE INTEGRATION

For a traditional image sensor, the response function between photon flux (measured by photons per sec) and the incident photons is linear and can be modeled as:

$$\Phi_{\text{cc}}(t) = \frac{1}{q_{\text{cc}}T} \int_{f-T/2}^{f+T/2} N(t) dt$$

where $N$ is the number of photons detected by the sensor, $q_{\text{cc}}$ is the quantum efficiency of the sensor and $T$ is the exposure time. During the exposure time, the latent image sequence $N(t)$ is expressed as a function of intensity changes $E(t)$ and the previous latent image $N(f)$ as follows:

$$N(t) = N(f) \exp\left(cE(t)\right) \qquad (2)$$

where $c$ is a constant threshold and $E(t)$ is the intensity change which is the integral of the event signal $e(t)$.

To deblur this image using events, the following event-based double integral (EDI) model was proposed in [9]:

$$\Phi_{\text{cc}}(t) = \frac{N(f)}{q_{\text{cc}}T} \int_{f-T/2}^{f+T/2} \exp\left(cE(t)\right) dt. \qquad (3)$$

## 4 COMPARISON TO QBP

We show qualitative results comparing the effect of bandwidth on QBP [10] and our method in the fan sequence in Fig. 3. For the same SPAD bandwidth, our method has less noise compared to QBP as we use events to deblur the images, on the other hand QBP uses binary frames to deblur. Moreover for the lower bandwidth, the binary frames are not sufficient for deblurring which results in noisier reconstruction.

## 5 QUALITATIVE RESULTS

We show more qualitative results on the real-sensor data HS-ESPAD in Fig. 4 and simulation data SimSPAD in Fig. 5. The event-baseline E2VID [6] is able to capture details in high contrast scenes as events are stronger in this region. However, it suffers to capture details in the low contrast region, where SPAD data is better. Thus our method combines the advantages of the two resulting in the best quality of image reconstruction.

## REFERENCES

[1] R. Graça, B. McReynolds, and T. Delbruck, "Shining light on the dvs pixel: A tutorial and discussion about biasing and optimization," in *Proceedings of the IEEE/CVF Conference on Computer Vision and Pattern Recognition (CVPR) Workshops*, June 2023, pp. 4045–4053.

[2] Q. Gao, X. Sun, Z. Yu, and X. Chen, "Understanding and controlling the sensitivity of event cameras in responding to static objects," in *2023 IEEE/ASME International Conference on Advanced Intelligent Mechatronics (AIM)*, 2023, pp. 783–786.

[3] C. Posch, D. Matolin, and R. Wohlgenannt, "A QVGA 143 dB dynamic range frame-free PWM image sensor with lossless pixel-level video compression and time-domain CDS," vol. 46, no. 1, pp. 259–275, Jan. 2011.

[4] A. Ingle, T. Seets, M. Buttafava, S. Gupta, A. Tosi, M. Gupta, and A. Velten, "Passive inter-photon imaging," in *Proceedings of the IEEE/CVF Conference on Computer Vision and Pattern Recognition (CVPR)*, June 2021, pp. 8585–8595.

[5] C. Scheerlinck, N. Barnes, and R. Mahony, "Continuous-time intensity estimation using event cameras," in *Computer Vision – ACCV 2018*, C. Jawahar, H. Li, G. Mori, and K. Schindler, Eds. Cham: Springer International Publishing, 2019, pp. 308–324.

[6] H. Rebecq, R. Ranftl, V. Koltun, and D. Scaramuzza, "High speed and high dynamic range video with an event camera," *IEEE Transactions on Pattern Analysis and Machine Intelligence*, vol. 43, pp. 1964–1980, 6 2021.

[7] X. Chu, L. Chen, and W. Yu, "Nafssr: Stereo image super-resolution using nafnet," in *Proceedings of the IEEE/CVF Conference on Computer Vision and Pattern Recognition (CVPR) Workshops*, June 2022, pp. 1239–1248.

[8] Z. Wang, Y. Ng, C. Scheerlinck, and R. Mahony, "An asynchronous kalman filter for hybrid event cameras," in *Proceedings of the IEEE/CVF International Conference on Computer Vision*, 2021, pp. 448–457.

[9] L. Pan, C. Scheerlinck, X. Yu, R. Hartley, M. Liu, and Y. Dai, "Bringing a blurry frame alive at high frame-rate with an event camera," in *Proceedings of the IEEE Conference on Computer Vision and Pattern Recognition*, 2019, pp. 6820–6829.

[10] S. Ma, S. Gupta, A. C. Ulku, C. Bruschini, E. Charbon, and M. Gupta, "Quanta burst photography," *ACM Trans. Graph.*, vol. 39, no. 4, aug 2020.

[11] H. Rebecq, G. Gallego, E. Mueggler, and D. Scaramuzza, "Emvs: Event-based multi-view stereo—3d reconstruction with an event camera in real-time," *International Journal of Computer Vision*, vol. 126, pp. 1394–1414, 12 2018.

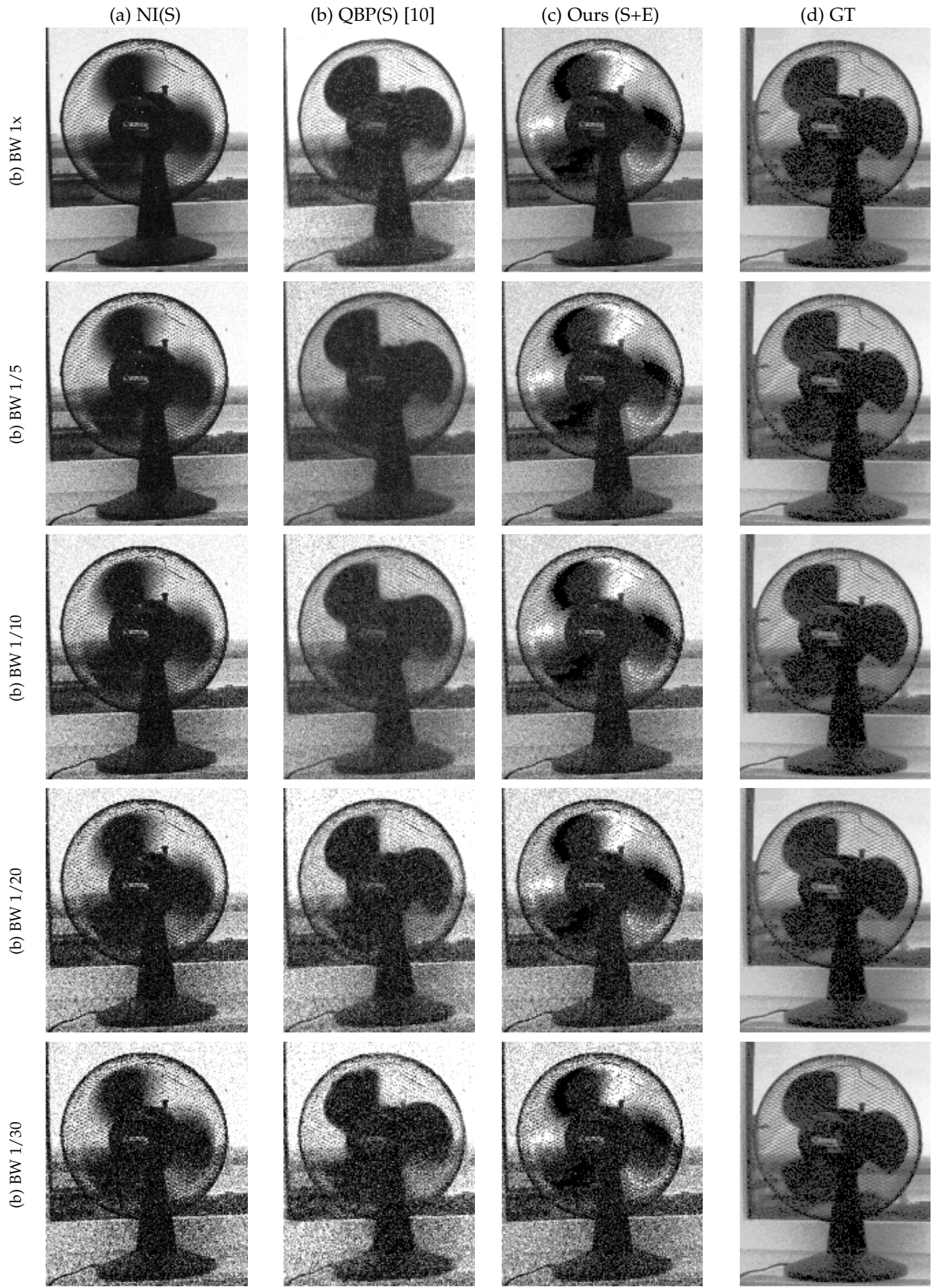

Fig. 3. Effect of bandwidth for (a) Naive integration (NI), (b) QBP [10] and (c) our method. Decreasing the bandwidth adds more noise to the images, however since we use events to deblur the images, our approach results in less noisy images

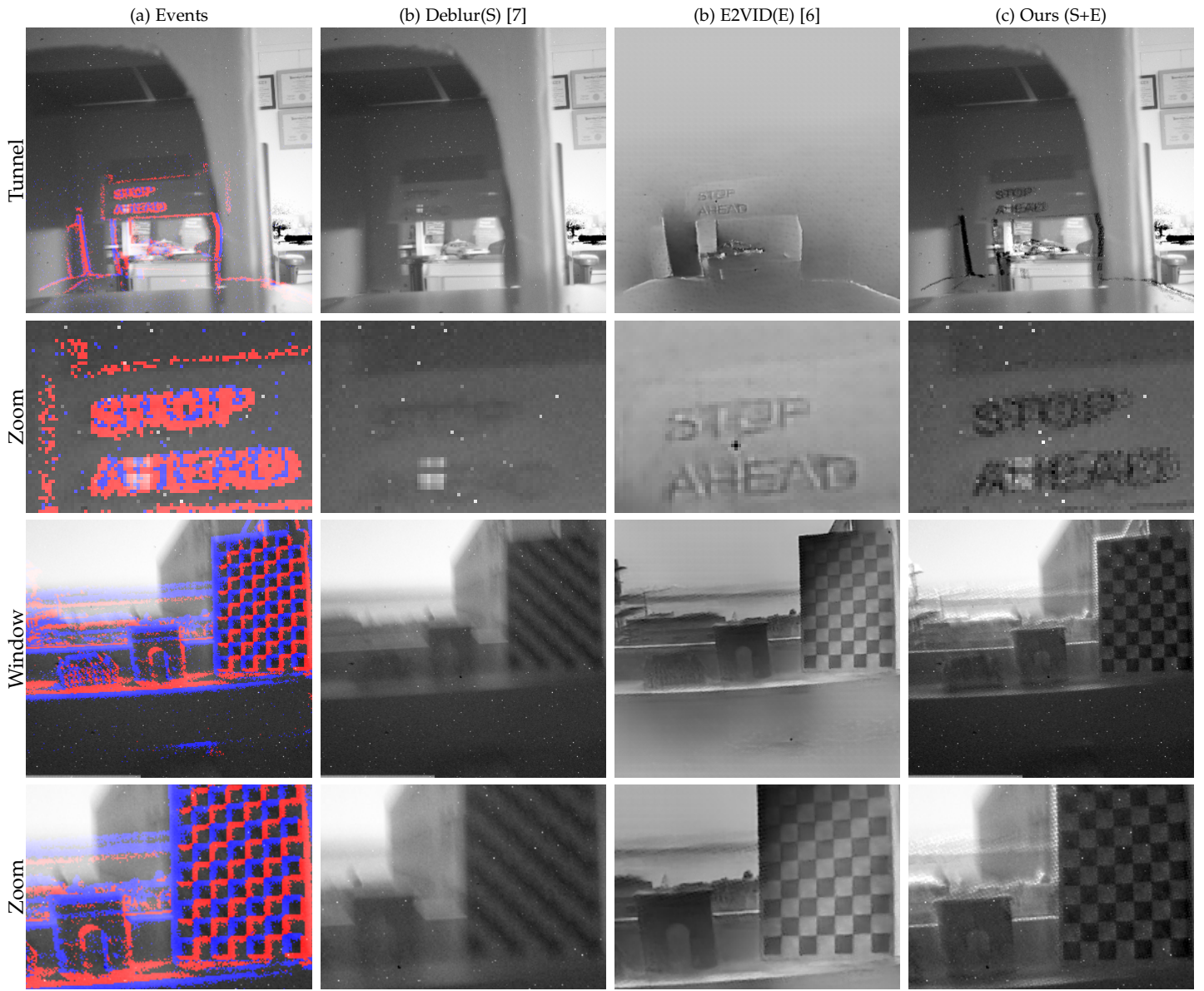

Fig. 4. Qualitative results comparing the best SPAD-only baseline [7] (b), event-only baseline $E2VID$ [6] (c) and our method (c) on HS-ESPAD. The aligned and synchronized events are overlaid on SPADs images and visualized in (a).

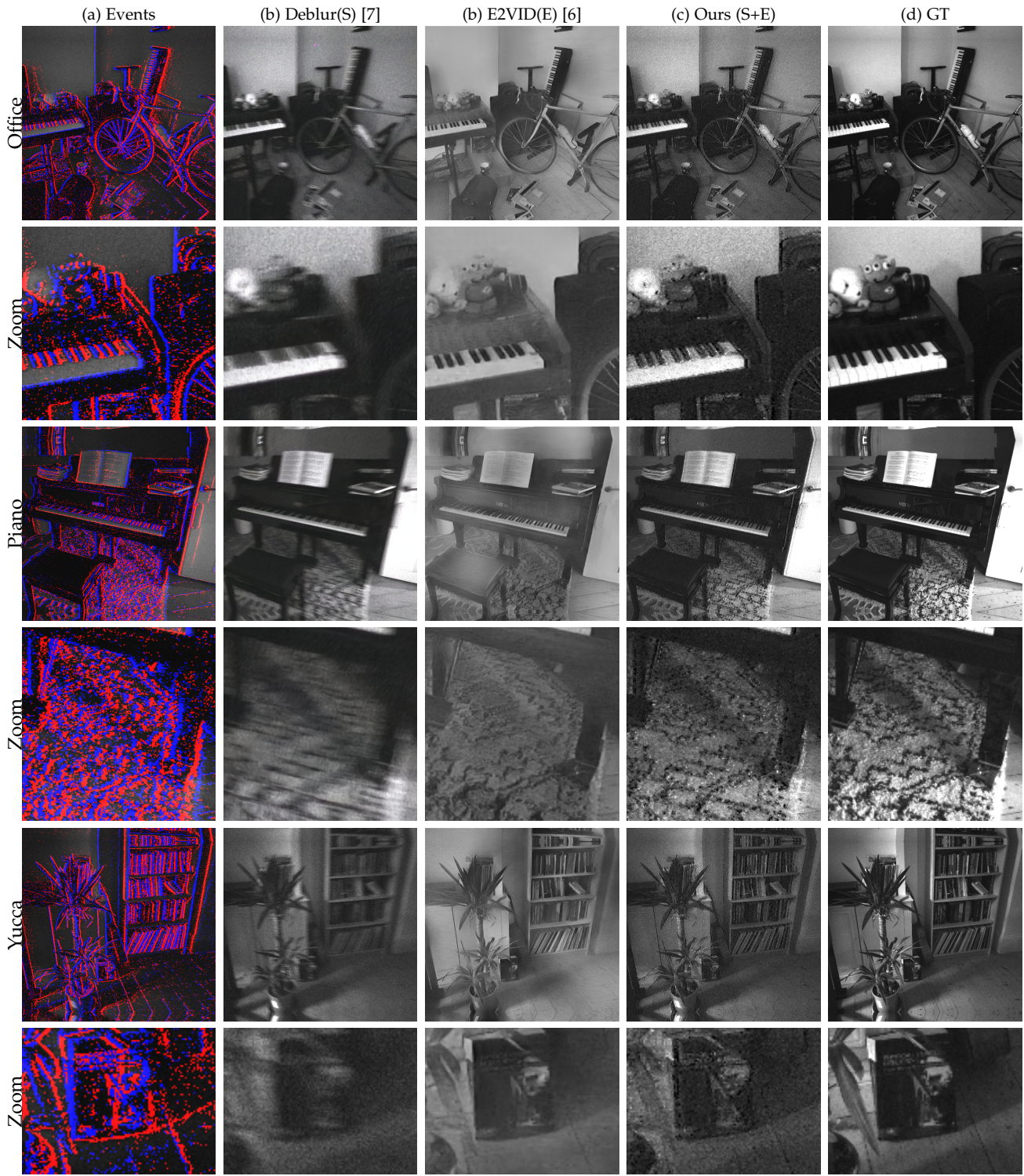

Fig. 5. Results of the SPAD-only baseline Deblur [7] (b), best Event-only baseline [11] (c) and our method (c) on our SimSPAD dataset. Events are overlaid on the blurred image (a).


\end{document}